\documentclass[journal,10pt]{IEEEtran}

\usepackage{amsthm,amssymb,graphicx,multirow,amsmath,color,amsfonts}
\usepackage[update,prepend]{epstopdf}
\usepackage{cite}
\usepackage[latin1]{inputenc}
\usepackage{tikz}
\usetikzlibrary{arrows,calc}		
\usepackage{bbm} 
\usepackage{pdfpages}
\usepackage{tabulary}
\usepackage{multirow}
\usepackage{comment}
\usepackage{breqn} 
\usepackage{subfig} 
\usepackage{bigints}

\usepackage{graphicx,xcolor,colortbl} 
\usepackage{caption} 

\def\sir{\mathtt{SIR}}

\theoremstyle{definition}
\newtheorem{defn}{Definition}

\newtheorem{theorem}{Theorem}

\newtheorem{lemma}{Lemma}

\usepackage{hyperref}

\usepackage{blindtext} 

\def\chr#1{{\color{black}#1}}

\include{notation}
\allowdisplaybreaks 

\begin{document}
	\graphicspath{{./Figures/}}
	\title{Delay Analysis in Full-Duplex Heterogeneous Cellular Networks}

\author{ Leila Marandi, Mansour Naslcheraghi, Seyed Ali Ghorashi, Mohammad Shikh-Bahaei
	
	\thanks{Copyright (c) 2015 IEEE. Personal use of this material is permitted. However, permission to use this material for any other purposes must be obtained from the IEEE by sending a request to pubs-permissions@ieee.org.}
	
	\thanks{L. Marandi and S. A. ghorashi are with Cognitive Telecommunication Research Group, Department of Telecommunications, Faculty of Electrical Engineering, Shahid Beheshti University, G.C., Tehran, Iran (email: le.marandi@gmail.com, a\textunderscore ghorashi@sbu.ac.ir), M. Naslcheraghi is with Department of Electrical Engineering, Polytechnique Montreal, Montreal, Quebec, Canada (email: mansour.naslcheraghi@polymtl.ca), and M. Shikh-Bahaei is with Department of Informatics, King's College London, London, UK (email: m.sbahaei@kcl.ac.uk).} 
}

\markboth{IEEE Transactions on Vehicular Technology,~Vol.~XX, No.~XX, XXX~2019}
{}

\maketitle

\bstctlcite{IEEEexample:BSTcontrol} 

\begin{abstract}
Heterogeneous networks (HetNets) as  a combination of macro cells and small cells are used to increase the cellular network's capacity, and present a perfect solution for high-speed communications. Increasing area spectrum efficiency and capacity of HetNets largely depends on the high speed of backhaul links. One effective way which is currently utilized in HetNets is the use of full-duplex (FD) technology that potentially doubles the spectral efficiency without the need for additional spectrum. On the other hand, one of the most critical network design requirements is delay, which is a key representation of the quality of service (QoS) in modern cellular networks. In this paper, by utilizing tools from the stochastic geometry, we analyze the local delay for downlink (DL) channel, which is
typically defined as the mean number of required time slots for a
successful communication. Given imperfect self-interference (SI) cancellation in practical FD communications, we utilize duplex mode (half-duplex (HD) or FD) for each user based on the distance from its serving base station (BS). Further, we aim to investigate the energy efficiency (EE) for both duplexing modes, i.e., HD and FD, by considering local delay. We conduct extensive simulations to validate system performance in terms of local delay versus different system key parameters.
\end{abstract}

\begin{IEEEkeywords}
	Heterogeneous Networks, Full-Duplex, Half-Duplex, Local Delay, Stochastic Geometry, Energy Efficiency.
\end{IEEEkeywords}

\IEEEpeerreviewmaketitle

\section{Introduction}

Heterogeneous networks (HetNets) provide services across both macro and small cells to optimize the mix of capabilities. As mobile broadband demand continues to grow dramatically, significant increases in capacity can be achieved by adding small cells to complement macro networks, forming HetNets. By the emergence of numerous applications in wireless networks, delay is one of the most important design measures of these networks that influences the quality of service (QoS) and accordingly the quality of experience (QoE) which directly affects the system reliability. Therefore, experts urge mobile operators to redesign their networks in more advanced ways to increase network capacity and support the growth of network traffic. Many technologies are introduced and developed in the fifth generation (5G) networks such as full-duplex (FD) communications, mmWave, Edge-Computing, Edge-Caching, etc. Meanwhile, HetNets consist of several layers of cells, generally operating at the same bandwidth. Accordingly, low power nodes (LPNs) including micro, pico, and femto BSs are distributed throughout the macro cells. In HetNets, macro cells provide widespread coverage, while LPNs use to reduce dead zones and more importantly to cover high traffic areas \cite{HetNet2} to improve QoS and overall network capacity. In half-duplex (HD) communications, the mutual connection between pairs is obtained by two methods of Frequency Division Duplexing (FDD) or Time Division Duplexing (TDD). This partitioning of resources in time or frequency domain restricts the use of resources in the network and such methods do not guarantee the 5G requirements \cite{HD}. One of the ways recently used in HetNets is exploiting FD radios in which nodes are capable to transmit and receive simultaneously in the same frequency and time. Recent developments in dealing with the self-Interference (SI) at FD radios promises practical realization of this technology \cite{FD1,FD2}. Recent studies on FD communication \cite{FD-app1,FD-app2,FD-app3,FDSurvey} reveal that the performance of FD outperforms its HD counterpart in terms of delay and sum throughput. \chr{Three important criteria for delivery of network information are throughput, delay and reliability\cite{LocalDelay1}.} Given delay requirements as an important design goal of the HetNets, it is not clear how system performs in terms of delay when the system is FD-capable. Also, since each tier in HetNet has distinctive characteristics, it significantly complicates the analysis as well as simulations. \chr{In the downlink, transmission to multiple users can be enabled within the same time-frequency resources \cite{EditorPaper} by applying superposition transmission and power control at the transmitter and successive interference cancellation (SIC) at the receiver.} We are also inspired by the challenge of both FD and HetNet techniques that require high power consumption, which is in contrast with the promise of green evolution for future communication systems. Detail of these challenges is still unclear, which is the main focus of this work. Recently, stochastic geometry has been used to study delay in wireless cellular networks. This study is more crucial in two types of delay; the queueing delay and the transmission delay \cite{Delay-cause}. Delay of the queue mainly refers to the waiting time in the service queue while the transmission delay is the time for a successful transmission of data. Recent studies \cite{LocalDelay1,LocalDelay2,LocalDelay3} introduced the concept of local delay that is a basic form of transmission delay. 

\chr{In this paper, to study the delivery of information in HetNet with FD capability, the local delay is considered. Local delay can guarantee reliability and throughput of the system, by acquiring the mean number of time slots for successful transmission. For some network parameters, this local delay will is very high, which means that for a successful transmission, we need a very high number of time slots\cite{LocalDelay1}. Therefore, evaluating local delay is significantly important to make sure that users are being served, reasonably. } The follow-up papers utilized this idea to analyze local delay in a clustered network \cite{LocalDelay4}. \chr{The authors in \cite{Reviewers,Reviewers2} introduced several approaches to match the various traffic arrival in ultra-dense networks (UDN). They utilized tools from the stochastic geometry to study the tradeoff between delay and security of the physical layer in wireless networks.} The authors in \cite{Hybrid-dulex1,Hybrid-duplex2,Hybrid-duplex3} proposed a hybrid-duplex scheme for HetNets by utilizing tools from the stochastic geometry and analyzed signal-to-interference-plus-noise ratio (SINR) distribution, area spectral efficiency (ASE), achievable rate and throughput for both DL and Uplink (UL) channels and have shown that Hybrid-duplex scheme outperforms both pure-HD and pure-FD HetNets. In another research direction \cite{LocalDelay5,DTX-LocalDelay, EE-HD}, the investigations of energy efficiency (EE) and local delay are conducted in conventional HetNets by considering HD mode for all links. However, these methods are inefficient for FD/Hybrid systems due to the SI and suffered interference from the FD users on the receiver of interest.


\subsection{Main Contributions and Outcomes}
Thus far, the existing literature have not investigated the concept of local delay in cellular HetNets with hybrid HD/FD communications. Given imperfect SIC in practical FD transceivers and possible aggregate interference from the other transmitters at the receiver of interest, it is unlikely to choose FD communication as the optimal mode for all nodes. Hence, depending on the random circumstances (e.g. distance, channel, etc.) in a cellular HetNet, one duplexing mode will be optimal. In this paper, we analyze local delay by exploiting distance-based mode selection scheme to choose optimal mode for network nodes. Imperfect SI cancellation in practical FD-enabled systems and massive deployment of small cells in HetNets not only increases the expected interference on the receiver of interest, but also causes more energy consumption. Thereby, we investigate energy efficiency (EE) in the FD-enabled HetNets which is eventually more than its HD counterpart. Thus far, from the local delay point of view, to the best of our knowledge, there is no existing work to investigate EE in the FD-Enabled HetNets. Given FD capability for all nodes in the \chr{network}, we aim to propose a hybrid duplex mode selection for the HetNets. Further details on the contributions are provided in the sequel.

\begin{itemize}
	\item We analyze the local delay for HetNets with Hybrid duplexing (HD/FD) and evaluate the impact of FD on the local delay by utilizing tools from the stochastic geometry. In particular, we aim to model and analyze the expected time required for a successful transmission from an arbitrary node to its respective receiver in terms of system key parameters.

	\item  We exploit a hybrid duplex mode selection  scheme for FD-enabled HetNets in which all BSs and user devices are FD-capable. \chr{The reason behind choosing the distance-based duplex mode selection policy is that, considering imperfect self-interference cancellation (SIC), the SIR of a full-duplex user may be worse than the half-duplex one in the same place. Although the small-scale fading affects the SIR,  it could be averaged out and the path-loss would be the main factor influencing the SIR in long term. Hence, it may not be suitable for the cell edge users to establish a full-duplex communication. Based on that and similar to \cite{Hybrid-duplex2, distance-criteria}, distance from the serving BS is chosen as a duplex mode selection criterion. 
	} This scheme is based on user distance from its respective serving BS. Different from \cite{Hybrid-dulex1, Hybrid-duplex2}, we consider silent mode capability for all BSs to reduce the energy consumption. This capability allows BSs to be inactive and this results in reduction of the experienced interference on the receiver of interest and correspondingly, will lead to increase the chance of successful transmissions.

	\item Different from \cite{EE-Hybrid} in which the EE is investigated in single layer cellular networks (none-HetNet), we investigate the EE in the FD-capable HetNets by considering local delay in the analysis and utilizing tools from the stochastic geometry. In particular, we have modeled EE with respect to the behavior of local delay.
\end{itemize}
The rest of this paper is organized as follows. In section \ref{sec:SysMod}, we delineate the system model and introduce the key parameters. In section \ref{sec:analysis LD}, we model and analyze the local delay. Theoretical and simulation results are provided in section \ref{sec:NumResults} and finally section \ref{sec:conclusion} concludes the paper.

\section{System Model} \label{sec:SysMod}
\subsection{Network Model} \label{sub: NetMod}
We consider a HetNet with $K-$ tier of BSs that are different in transmitting powers, densities and path loss exponents. The BS locations of tier $\mathcal{K} = \left\{ {1,2, \cdots ,K} \right\}$ are distributed in the Euclidean plane $\mathbb{R}^2$ independently based on the homogeneous Poisson point process (HPPP) denoted by ${\Phi _k}$ with density of ${\lambda _k}$. The BSs of each tier have the same transmit power, which are denoted by ${p_k}$. Due to high density of BSs in HetNets, the background noise is negligible in comparison with the amount of interference. \chr{Denoting $\tau_k$ as the system target signal-to-interference ratio (SIR) threshold for a typical user located in tier $k$ and is operating in mode $\dag$ (FD/HD), the probability of the event that this typical user satisfies system threshold $\tau_k$ is defined as $\mathcal{P}\left( {\textup{SIR}_k^{\dag} \ge {\tau _k}} \right)$. If the SIR thresholds of all tiers are the same, then we have $\tau_k = \tau$.} In general, the properties of each tier can be expressed as $\left\{{{\lambda _k},{{\text{p}}_k},{\tau _k}} \right\}$. Users are also independently distributed in the Euclidean plan based on HPPP ${\Phi _u}$ with density of ${\lambda _u}$. The transmit power of users is denoted by ${p_u}$. Both BSs and users are FD-capable. Imperfect SI cancellation is assumed with the residual SI signal power of $\beta {p_k}$, $\beta {p_u}$, for BSs and users, respectively, where $0 \le \beta  \le 1$ represents the residual power ratio. If $\beta = 0$, the SI cancellation is perfect and in case of $\beta = 1$, there is no SI cancellation. The network is considered to be static, i.e., the positions of BSs and users remain the same throughout time. \chr{In practical systems, the delay performance highly relies on the properties of the traffic in wireless networks. In this paper, to consider the traffic modeling, the scenario of the network is considered as backlogged \cite{Reviewers2}, i.e., when the transmitters are scheduled to communicate, they always have data to transmit.} Furthermore, we assume the system to be open access, where the user can connect to any BS. Inspiring from the Slivniak's theorem \cite{typicaluser-Palm}, we conduct the analysis on a typical user in the origin for DL channel, for this purpose, which implies that any user has the same chance of being at the origin in process. \chr{The notations of the paper are summarized in table \ref{table:Notation}.}
\vspace*{-3mm}

\subsection{Channel Model and Duplex mode selection} \label{sub:Channel Model Subsection}
In the proposed model, each BS and user have a single antenna for power transmission/reception. For the propagation model, we assume a combination of fading and path loss. It is assumed that all channels are Rayleigh fading with unit mean. Let us denote the power fading coefficient between BS $b_k$ and a typical user $u_o$ at given time slot $t$ with ${h_{t,{b_k}{u_o}}} \sim \exp \left( 1 \right)$. We assume standard power-law path loss model $l\left( r \right) = {\left\| r \right\|^{ - {\alpha _k}}}$, where the BS of tier $k$ is located at distance $r$ from the origin (typical user) and $\alpha  > 2$ is the path loss exponent. We consider cell association based on maximum average DL received power (DRP), where a mobile user is associated with the strongest BS in terms of long-term averaged DRP at the user. The tier index that a typical user is associated to, is given by
\begin{equation} \label{eq: maximum DRP}
k = \arg \mathop {\max }\limits_i \left\{ {{p_i}{{\left\| {{r_i}} \right\|}^{ - {\alpha _i}}},i \in \mathcal{K}} \right\}.
\end{equation}

By exploiting distance-based duplex mode selection, we denote ${\vartheta _k}$ as the distance threshold; if the distance between typical user and its serving BS is less than ${\vartheta _k}$, the user will be treated as an FD user, otherwise, it will operate in HD mode. FD and HD user sets in tier $k$ are denoted by  ${\mathcal{U}_{{k_{\rm FD}}}}$, ${\mathcal{U}_{{k_{\rm HD}}}}$, respectively. The corresponding definitions can be written as
\begin{equation}\label{eq: set of FD user }
{\mathcal{U}_{{k_{\rm FD}}}} = \left\{ {\left( {{p_k}{{\left\| {{r_k}} \right\|}^{ - {\alpha _k}}} > {p_j}{{\left\| {{r_j}} \right\|}^{ - {\alpha _j}}},\forall j \ne k} \right) \wedge \left( {{r_k} \leqslant {\vartheta _k}} \right)} \right\},
\end{equation}
\begin{equation}\label{eq: set of HD user }
{\mathcal{U}_{{k_{\rm HD}}}} = \left\{ {\left( {{p_k}{{\left\| {{r_k}} \right\|}^{ - {\alpha _k}}} > {p_j}{{\left\| {{r_j}} \right\|}^{ - {\alpha _j}}},\forall j \ne k} \right) \wedge \left( {{r_k} > {\vartheta _k}} \right)} \right\}.
\end{equation}

Time is divided into discrete slots with the same duration ${\delta _t}$ and each transmission spans one time slot. BSs can transmit/receive data only if they are in the active mode. Technically, each BS of the ${k^{th}}$ tier at any time slot with probability ${\chi_{_k}}\left( {0 \le {\chi_{_k}} \le 1} \right)$ is silent. The interference at typical user in time slot $t$ consists of interference from all active BSs, except the serving BS, as well as the set of FD users, which can be expressed as
\begin{equation}\label{eq: interfernce in time slot}
\begin{split}
{{\mathcal{I}}_t} = \sum\limits_{j \in \mathcal{K}}\Bigg[& \sum\limits_{{b_j} \in \Phi _{ja}\backslash \left\{ {{b_o}} \right\}} {{p_j}{h_{t,{b_j}{u_o}}}{{\left\| {{r_{{b_j}{u_o}}}} \right\|}^{ - {\alpha _j}}} + }\\
&\sum\limits_{{u_i} \in \Phi _u^{\rm FD}\backslash \left\{ {{u_o}} \right\}} {{p_u}{h_{t,{u_i}{u_o}}}{{\left\| {{r_{{u_i}{u_o}}}} \right\|}^{ - {\alpha _u}}}}  \Bigg], 
\end{split}
\end{equation} 
where ${\Phi _{ja}} = \Phi _{ja}^{\rm FD} \cup \Phi _{ja}^{\rm HD}$ indicate the set of active BSs that consists of active FD $\left( {\Phi _{ja}^{\rm FD}}\right)$ and HD $\left( {\Phi _{ja}^{\rm HD}}\right)$ BSs in tier $j$ in time slot $t$. The interference caused by the active BSs can be written separately; ${{\mathcal{I}}_{t,B}} = \sum\nolimits_{{b_j} \in \Phi _{_{ja}}^{\rm FD}\backslash \left\{ {{b_o}} \right\}} {{p_j}{h_{t,{b_j}{u_o}}}{{\left\| {{r_{{b_j}{u_o}}}} \right\|}^{ - {\alpha _j}}}}  + \sum\nolimits_{{b_j} \in \Phi _{_{ja}}^{\rm HD}\backslash \left\{ {{b_o}} \right\}} {{p_j}{h_{t,{b_j}{u_o}}}{{\left\| {{r_{{b_j}{u_o}}}} \right\|}^{ - {\alpha _j}}}} $. $\Phi _{u}^{\rm FD}$ is the set of FD users. ${b_o}$ represents serving BS of typical user ${u_o}$. ${{h_{t,{b_j}{u_o}}}}$ indicates the fading power coefficients between BS ${b_j}$ and typical user ${u_o}$, similarly ${{h_{t,{u_i}{u_o}}}}$ is fading from FD user ${u_i}$ and typical user ${u_o}$. The SIR of the typical user in time slot $t$ in the ${k^{th}}$ tier is 
\begin{equation} \label{eq: SIR FD}
{\sir} _{k,t}^{{\dag} } = \frac{{{p_k}{h_{t,{b_o}{u_o}}}{{\left\| {{r_{{b_o}{u_o}}}} \right\|}^{ - {\alpha _k}}}}}{{{{\mathcal{I}}_t} + { \mathbbm{1}_\dag}. \beta {p_{{u_o}}}}},
\end{equation}
where $\dag \in \{ \rm HD, FD  \}$. ${\sir}_{k,t}^{\rm FD}$, ${\sir}_{k,t}^{\rm HD}$ indicate the SIR, when typical user is FD and HD user, respectively. ${\beta {p_{{u_o}}}}$ is residual SI signal at the typical user. $\mathbbm{1}_\dag$ is the indicator function and defined by
\begin{equation}
\label{IndFunc}
{\mathbbm{1}_\dag} = \left\{ {\begin{array}{*{20}{c}}
	1 ,&{{\dag} = \textup{FD}},\\
	0,&{\dag = \textup{HD}}.
	\end{array}} \right.
\end{equation}

\begin{figure}
	\centering
	\includegraphics[width=0.35
	\textwidth]{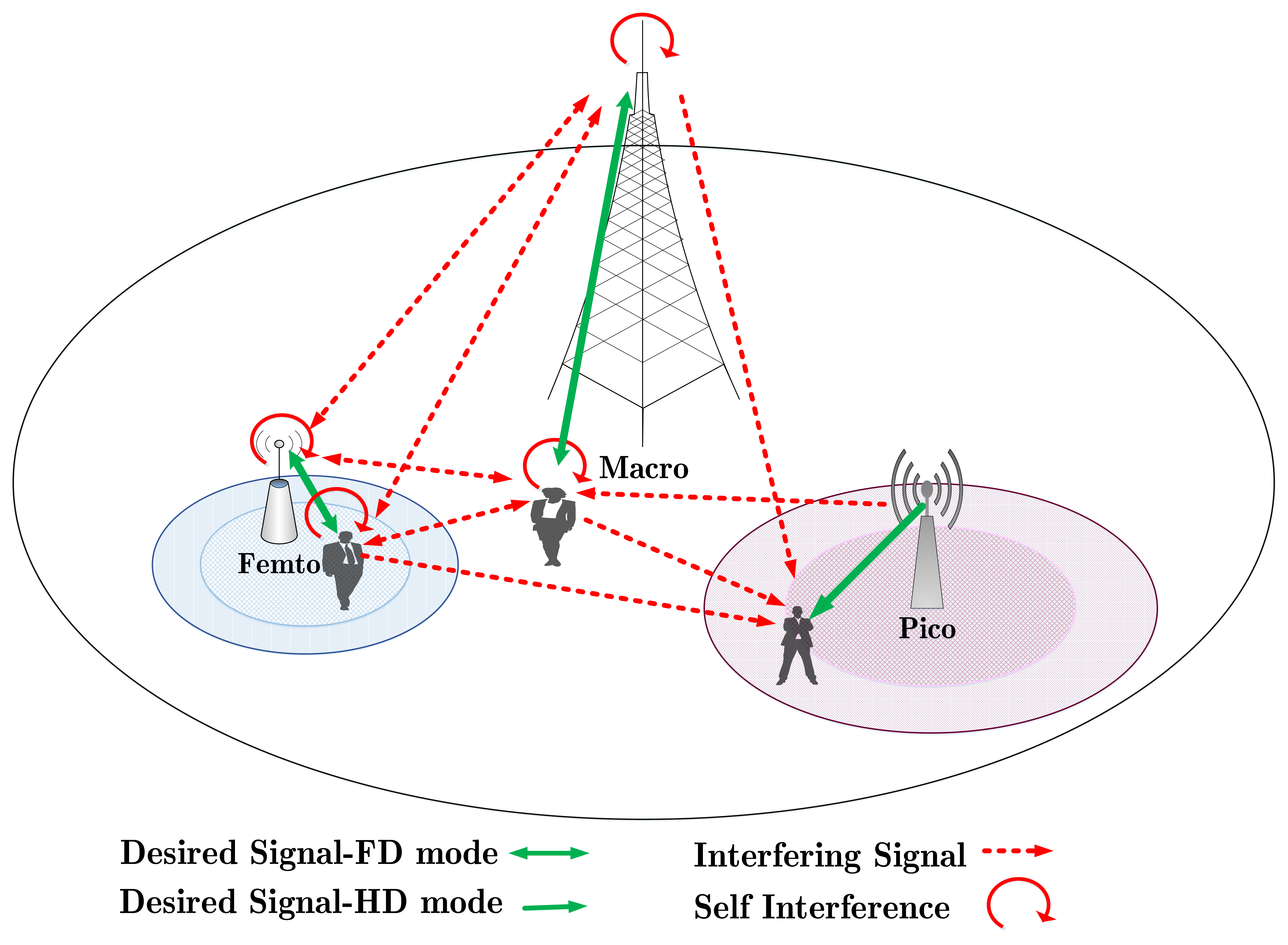}
	\caption{\scriptsize An illustration of three-tier HetNet with hybrid HD/FD duplexing modes along with the desired and interfering signals.}
	\label{fig: Sys Model}
\end{figure}

An illustration of the system described above is delineated for a 3-tier HetNet in Fig.\ref{fig: Sys Model} along with the expected interferences in DL channel. Tier 1, i.e. Macro has more transmission power than Pico (i.e. tier 2) and Femto (i.e. tier 3) cells. Shown HD/FD modes for the users are determined based on distance as described before.

\arrayrulecolor{black}
\begin{table}
	\captionsetup{font={scriptsize, color=black}}
	\begin{center}
		\caption{\MakeUppercase{SUMMARY OF NOTATION}}
		\label{table:Notation}
		\begin{tabular}{|>{\color{black}}c||>{\color{black}}c|}
			\hline	
			\textbf{Notation} & \textbf{Description}\\ \hline \hline
			${\Phi _k}, {\Phi _u}$ &
			\begin{tabular}{c}
				Point Process of BSs in tier $k$/\\ Point Process of users
			\end{tabular}\\ \hline	 
			${\Phi _{ja}}, \Phi _{ja}^{\rm FD}, \Phi _{ja}^{\rm HD}$ &
			\begin{tabular}{c}
				Point Process of active FD BSs/ \\Point Process of active HD BSs in tier $j$
			\end{tabular}  \\ \hline			
			${\lambda _k}, {\lambda _u}$ &
			\begin{tabular}{c}
				Density of BSs in tier $k$/\\ density of users 
			\end{tabular}\\ \hline
			${{p_k}}, {{p_u}}$ & 
			\begin{tabular}{c}
				Transmit power of BSs in tier $k$/ \\transmit power of users
			\end{tabular} \\ \hline		
			${{\tau _k}}$& SIR threshold of tier $k$ \\ \hline
			${{\chi _k}}$ & Silent (sleep) probability in tier $k$ \\ \hline
			${{r_k}}$ &
			\begin{tabular}{c} 
				Distance from the typical user to target BS in tier $k$ 
			\end{tabular} \\ \hline
			$\alpha$ & Path loss exponent\\ \hline
			${h_{t,{b_k}{u_o}}}, {h_{t,{u_i}{u_o}}}$ &
			\begin{tabular}{c}
				Power fading between BS $b_k$ and a typical user $u_o$/\\ power fading from FD user ${u_i}$ and typical user ${u_o}$\\ at given time slot $t$
			\end{tabular} \\ \hline
			$\beta {p_k}, \beta {p_u}$ & 
			\begin{tabular}{c}
				Residual SI signal power of the BSs in tier $k$/\\ Residual SI signal power of users
			\end{tabular}\\ \hline
			${\vartheta _k}$ & Distance threshold\\ \hline
			$\mathcal{U}_{{k_{\rm FD}}}, \mathcal{U}_{{k_{\rm HD}}}$ & FD user sets/ HD user sets in tier $k$ \\ \hline	
			$\mathcal{A}_k, {\mathcal{A}_k^{\rm FD}}, {\mathcal{A}_k^{\rm HD}}$ & Association probability of the typical user to tier $k$\\ \hline
			${{\mathcal{I}}_t}$	 & Interference at typical user in time slot $t$  \\ \hline 		
		\end{tabular}
	\end{center}
\end{table}

\section{Local Delay Analysis} \label{sec:analysis LD}
First we derive general local delay on DL channel for a typical operating in \chr{FD and HD} modes, then we calculate EE based on the local delay analysis. 

\chr{\subsection{Local Delay Definition}}
	\chr{Let $r_k$ be the distance between the typical user and its associated BS. Assuming the typical user at the origin, we assume that $\mathcal{P}_{\textup{suc}}^{\dag}$ is the probability that the typical user (operating in mode $\dag$) is successfully connected to its tagged BS in a single transmission conditioned on $r_k$. Then, the conditional success probability is given by}

\chr{\begin{equation} \label{eq: define suc prob}
	{{\mathcal{P}}_{\rm suc}^{\dag}} = \left( {1 - {\chi_{_k}}} \right){\mathcal{P}}\left( {{{\sir}_{k,t}^{{\dag} } } > \tau_k \left| {{r_k}} \right.} \right).
	\end{equation}}
\chr{where $\chi_k$ is the silence probability and $\tau_k$ is the given threshold defined in section \ref{sec:SysMod}. Successful transmission occurs only when the BS is active and SIR satisfies the system threshold value. The purpose of operating in silent mode is energy saving. However, if the associated user is going to transmit data and BS is in the silent mode, it should be noted that it is only for one time slot and not for the entire time duration, therefore, it can try to transmit in the next time slots. Conditioned on $r_k$, by considering the constant transmit power (or independent and identically distributed (i.i.d) transmit power) and i.i.d fading, the event of successful transmission in time slots and the success indicator random variables are temporally i.i.d \cite{LocalDelay6}. Hence, the distribution of the conditional local delay is geometric with mean $({{\mathcal{P}}_{\rm suc}})^{-1}$. The local delay of a typical user \cite{LocalDelay2,LocalDelay4} in tier $k$ operating in mode $\dag$ is then the expectation with respect to (w.r.t.) $r_k$:  }
\chr{\begin{equation}\label{eq: local delay of FD/HD typical user in tier k}
	\mathcal{D}_k^\dag  = {\mathbb{E}_{{r_k}}}\left[ {\frac{1}{{{\mathcal{P}}_{suc}^\dag }}} \right].
	\end{equation}}
\chr{In Eqn. \ref{eq: local delay of FD/HD typical user in tier k}, $\mathcal{D}_k^\dag$ denotes the average number of slots that takes the tagged BS to successfully transmit a packet to the typical user.
}
Note that the typical user associates to one tier. Now, according to the law of total probability, the local delay can be written as
\begin{equation} \label{eq: define total local delay FD/HD}
{\mathcal{D}_\dag } = \sum\limits_{k \in \mathcal{K}} {\mathcal{A}_k^\dag } \mathcal{D}_k^\dag,
\end{equation}

where ${\mathcal{A}_k^{\rm FD}}$ and $\mathcal{A}_k^{\rm HD}$ provide the per tier association probability that typical user operates in FD and HD modes, respectively. 

\begin{lemma} \label{lemma: association Prob.}
	Generally, the association probability in the ${k^{th}}$ tier for a typical user that operates in HD or FD mode is given by
\chr{\begin{equation} \label{eq: associate_FD prob}
	\begin{split}
	\resizebox{1\hsize}{!}{${\mathcal{A}_k^{\rm FD}} = 2\pi {\lambda _k}\bigintsss_0^{{\vartheta _k}} {x\exp \left( \sum\limits_{i \ne k} { - \pi {\lambda _i}{{\left( {\frac{{{p_i}}}{{{p_k}}}{x^{{\alpha _k}}}} \right)}^{\frac{2}{{{\alpha _i}}}}} - \pi {\lambda _k}{x^2}} \right)} dx.$}
	\end{split}
	\end{equation}}		
	and
	\chr{	\begin{equation} \label{eq: associate_HD prob}
		\begin{split}
		\resizebox{1\hsize}{!}{${\mathcal{A}_k^{\rm HD}} = 2\pi {\lambda _k}\bigintsss_{\vartheta _k}^{\infty } {x\exp \left( \sum\limits_{i \ne k} { - \pi {\lambda _i}{{\left( {\frac{{{p_i}}}{{{p_k}}}{x^{{\alpha _k}}}} \right)}^{\frac{2}{{{\alpha _i}}}}} - \pi {\lambda _k}{x^2}} \right)} dx.$}
		\end{split}
		\end{equation}}
 The probability that the user is associated to the tier $k$ is defined as the sum of the association probabilities in FD and HD modes, namely 
	\chr{\begin{equation} \label{eq: associate prob}
		\begin{split}
		\resizebox{1\hsize}{!}{${\mathcal{A}_k} = 2\pi {\lambda _k}\bigintsss_{0}^{\infty } {x\exp \left( \sum\limits_{i \ne k} { - \pi {\lambda _i}{{\left( {\frac{{{p_i}}}{{{p_k}}}{x^{{\alpha _k}}}} \right)}^{\frac{2}{{{\alpha _i}}}}} - \pi {\lambda _k}{x^2}} \right)} dx.$}
		\end{split}
		\end{equation}}
	
		\begin{IEEEproof} 
		see appendix \ref{appendix-Associate}.
	\end{IEEEproof}
\end{lemma}

\subsection{General Case of Local Delay} \label{sub: general result}
\begin{theorem} \label{theory: Local delay}
	The local delay in a $K$-tier HetNet with Hybrid duplex scheme and associated cell based on DRP is given by
\end{theorem}

\chr{
\begin{equation}\label{eq: General Eq local delay FD}
\begin{split}
{\mathcal{D}_{\rm FD}} = &\sum\limits_{k \in \mathcal{K} } {\frac{{2\pi {\lambda _k}}}{{\left( {1 - {\chi_{_k}}} \right)}}} \int_0^{{\vartheta _k}} {\exp \left( {s\beta {p_{{u_o}}}} \right)} {{\cal F}_1}\left( {r,\lambda_j } \right)\\
&\prod\limits_{j \in \mathcal{K}} {{{\cal F}_2}\left( {r,\lambda_j ,\tau_k } \right){{\cal F}_3}\left( {r,\lambda_u ,\tau_k } \right)} dr,
\end{split}	     
\end{equation}
}	
\chr{
\begin{equation}\label{eq: General Eq local delay HD}
\begin{split}
{\mathcal{D}_{\rm HD}} = & \sum\limits_{k \in \mathcal{K}} {\frac{{2\pi {\lambda _k}}}{{\left( {1 - {\chi_{_k}}} \right)}}} \int_{{\vartheta _k}}^\infty {{\cal F}_1}\left( {r,\lambda_j } \right) \prod\limits_{j \in \mathcal{K}} {{\cal F}_2}\left( {r,\lambda_j ,\tau_k } \right)\\
&{{\cal F}_3}\left( {r,\lambda_u ,\tau_k } \right) dr,
\end{split}
\end{equation}
}
where $s = \frac{{{\tau _k}r_k^{{\alpha _k}}}}{{{p_k}}}$,
\chr{\begin{equation}
\begin{split}
{{\cal F}_1}\left( {r,\lambda_j } \right) = r.\exp \left( { - \pi \sum\limits_{j \in \mathcal{K}} {{\lambda _j}{{\left( {\frac{{{p_j}}}{{{p_k}}}{r^{{\alpha _k}}}} \right)}^{\frac{2}{{{\alpha _j}}}}}} } \right),
\end{split}
\end{equation}}
\chr{\begin{equation}
	\begin{split}
	{{\cal F}_2}\left( {r,\lambda_j ,\tau_k } \right) = \exp \bigg( &\pi \left( {1 - {\chi_{_j}}} \right)\lambda _{_j}{{\left( {\frac{{{p_j}}}{{{p_k}}}{r^{{\alpha _k}}}} \right)}^{\frac{2}{{{\alpha _j}}}}}\\
	&\int_1^\infty  {\left( {\frac{{{\tau _k}}}{{{\tau _k} + {u^{\frac{{{\alpha _j}}}{2}}}}}} \right)du}\bigg),
	\end{split}
\end{equation}}
\chr{\begin{equation}
	\begin{split}
	{{\cal F}_3}\left( {r,\lambda_u ,\tau_k } \right) = \exp \left( {2\pi \lambda _{j,u}^{\rm FD}\int\limits_{{e_{k\left( j \right)}}} {\left( {1 - \frac{1}{{1 + s{p_u}{y^{ - {\alpha _u}}}}}} \right)ydy} } \right). \notag
	\end{split}
\end{equation}}

\begin{IEEEproof}
	We provide the proof for typical user in FD mode and similar to the same trend, it is valid for the typical user in HD mode. First, we need to get the probability of successful transmission by using Eq.(\ref{eq: define suc prob}). Then we put the final result of Eq.(\ref{eq: define suc prob}) in Eq.(\ref{eq: local delay of FD/HD typical user in tier k}) to obtain expression for the local delay when typical user operates in FD mode.
	\begin{IEEEeqnarray}{lLl}
		\label{eq: proof success prob}
		&{\mathcal{P}}_{\rm suc}^{\rm FD} = \left( {1 - {\chi_{_k}}} \right){\mathcal{P}}\left( {{\sir} _{k,t}^{\rm FD} > {\tau _k}\left| {{r_k}} \right.} \right)\nonumber\\
		&= \left( {1 - {\chi_{_k}}} \right){\mathcal{P}}\left( {{p_k}{h_{t,{b_o}{u_o}}}{{\left\| {{r_{{b_o}{u_o}}}} \right\|}^{ - {\alpha _k}}} > {\tau _k}\left( {{{\mathcal{I}}_t} + \beta {p_{{u_o}}}} \right)\left| {{r_k}} \right.} \right)\nonumber\\
		&= \left( {1 - {\chi_{_k}}} \right){\mathcal{P}}\left( {{h_{t,{b_o}{u_o}}} > \frac{{{\tau _k}{{\left\| {{r_{{b_o}{u_o}}}} \right\|}^{{\alpha _k}}}}}{{{p_k}}}\left( {{{\mathcal{I}}_t} + \beta {p_{{u_o}}}} \right)\left| {{r_k}} \right.} \right)\nonumber\\
		&\mathop  = \limits^a \left( {1 - {\chi_{_k}}} \right)\exp \left( { - s\beta {p_{{u_o}}}} \right){\mathbb{E}_{{{\mathcal{I}}_t}}}\left[ {\exp \left( { - s{{\mathcal{I}}_t}} \right)\left| {{r_k}} \right.} \right]\nonumber\\
		&= \left( {1 - {\chi_{_k}}} \right)\exp \left( { - s\beta {p_{{u_o}}}} \right){{\cal L}_{{{\mathcal{I}}_t}}}\left( {s\left| {{r_k}} \right.} \right),
	\end{IEEEeqnarray}
where (a) follows from the fact that ${h_{t,{b_o}{u_o}}} \sim \exp \left( 1 \right)$ and ${{\cal L}_{{{\mathcal{I}}_t}}}\left( {s\left| {{r_k}} \right.} \right)$ refers to the Laplace transform of interference and can be calculated as shown in Eq. (\ref{eq: expectation of interference}). where  ${\mathcal{I}_{t,j}} = {{\mathcal{I}}_{t,B}} + \sum\nolimits_{{u_i} \in \Phi _u^{FD}\backslash \left\{ {{u_o}} \right\}} {{p_u}{h_{t,{u_i}{u_o}}}{{\left\| {{r_{{u_i}{u_o}}}} \right\|}^{ - {\alpha _u}}}} $ and for ${\mathbb{E}_{\Phi ,h}}\left[ {\exp \left( { - s\sum\limits_{j \in \mathcal{K}} {{{\mathcal{I}}_{t,j}}} } \right)} \right]$, we get {Eq. (\ref{eq: expectation of interference})} where (a) is obtained by using probability generating functional (PGFL) of the Poisson point process (PPP). Limits of integrals {in Eq.(\ref{eq: expectation of interference}) } are from ${e_k}\left( j \right)$ tend to $\infty$ and therefore, the closest interferer from tier $j$ is located at least at the distance ${e_k}\left( j \right) = {\left( {\frac{{{p_j}}}{{{p_k}}}r_k^{{\alpha _k}}} \right)^{\frac{1}{{{\alpha _j}}}}}$. On the other hand, it is difficult to specify the exact range of interference from FD users to a typical user. Interfering users are distributed around their serving BSs, so it can be approximated by distance between interfering BSs and typical user, hence the limits of integrals are also ${e_k}\left( j \right)$. Since each BS specifies the duplex mode and how it connects to its user, the density of FD users is defined as $\lambda _{j,u}^{\rm FD} = \frac{{\mathcal{A}_j^{\rm FD}}}{{{\mathcal{A}_j}}}\left( {1 - {\chi_{_j}}} \right){\lambda _u}$, where ${{\mathcal{A}_j^{\rm FD}}}$ is the association probability that the user operates in FD mode within the ${j^{th}}$ tier. (b) is obtained by variables substitutions $u = {\left( {\frac{{{p_j}}}{{{p_k}}}r_k^{{\alpha _k}}} \right)^{ - \frac{2}{{{\alpha _j}}}}}{y^2}$ and finally, putting the Eq.(\ref{eq: proof success prob}) in Eq.(\ref{eq: local delay of FD/HD typical user in tier k}) gives the final result for $\mathcal{D}_k^{\rm FD}$. According to \cite{Stochastic-HetNet2, Hybrid-duplex2}, distance distribution of a typical user that operates in FD mode from its serving BS is given by

\begin{figure*}[t]  
	\chr{\begin{IEEEeqnarray}{lLl}\label{eq: expectation of interference}
  & {{\cal L}_{{{\mathcal{I}}_t}}}\left( s \right) = \mathbb{E}\left[ {\exp \left( { - s{{\mathcal{I}}_t}} \right)} \right] = {\mathbb{E}_{\Phi ,h}}\left[ {\exp \left( { - s\sum\limits_{j \in \mathcal{K}} {{{\mathcal{I}}_{t,j}}} } \right)} \right] 
		= \prod\limits_{j \in \mathcal{K} } {{\mathbb{E}_{\Phi ,h}}\left[ {\exp \left( { - s{{\mathcal{I}}_{t,j}}} \right)} \right]} \nonumber \\
		&= \prod\limits_{j \in \mathcal{K} }{\mathbb{E}_{\Phi ,h}} \Bigg[{\prod\limits_{{b_j} \in \Phi _{ja}\backslash \left\{ {{b_o}} \right\}}} \exp \left( { - s{p_j}{h_{t,{b_j}{u_o}}}{{\left\| {{r_{{b_j}{u_o}}}} \right\|}^{ - {\alpha _j}}}} \right) 
	    \prod\limits_{{u_i} \in \Phi _u^{\rm FD}\backslash \left\{ {{u_o}} \right\}} \exp \left( { - s{p_u}{h_{t,{u_i}{u_o}}}{{\left\| {{r_{{u_i}{u_o}}}} \right\|}^{ - {\alpha _u}}}} \right)\Bigg]\nonumber\\
		&\mathop  = \limits^{(a)} \prod\limits_{j \in \mathcal{K} } \Bigg(\exp \left(  - 2\pi \left( {1 - {\chi_{_j}}} \right)\lambda _{_j}\int_{{e_{k\left( j \right)}}}^\infty {\left( {1 - \frac{1}{{1 + s{p_j}{y^{ - {\alpha _j}}}}}} \right)ydy}  \right)
		\exp \left( { - 2\pi \lambda _{_j,u}^{\rm FD}\int_{{e_{k\left( j \right)}}}^\infty {\left( {1 - \frac{1}{{1 + s{p_u}{y^{ - {\alpha _u}}}}}} \right)ydy} } \right)\Bigg)\nonumber\\ 
		&\mathop  = \limits^{(b)} \prod\limits_{j \in \mathcal{K} } \Bigg(\exp \left( { - \pi \left( {1 - {\chi_{_j}}} \right)\lambda _{_j}{{\left( {\frac{{{p_j}}}{{{p_k}}}r_k^{{\alpha _k}}} \right)}^{\frac{2}{{{\alpha _j}}}}}\int_1^\infty  {\left( {\frac{{{\tau _k}}}{{{\tau _k} + {u^{\frac{{{\alpha _j}}}{2}}}}}} \right)du} } \right) \exp \left( { - 2\pi \lambda _{j,u}^{\rm FD}\int_{{e_{k\left( j \right)}}}^\infty {\left( {1 - \frac{1}{{1 + s{p_u}{y^{ - {\alpha _u}}}}}} \right)ydy} } \right)\Bigg).
	\end{IEEEeqnarray}}
	\hrule
\end{figure*}

\begin{equation}\label{eq: distance distribution FD}
{f_{r_k^{\rm FD}}}\left( r \right)= 
\begin{cases}
0&{r > {\vartheta _k,}}\\
{\frac{1}{{A_k^{\rm FD}}}2\pi {\lambda _k}r.\exp \left( { - \pi \sum\limits_{j \in \mathcal{K} } {{\lambda _j}{{\xi _k^2}\left( {{r}} \right)}} } \right)}&{r \le {\vartheta _k.}}
\end{cases}
\end{equation}

where ${\xi _k}\left( {{r}} \right) = {\left( {{r}^{{\alpha _k}}\frac{{{p_j}}}{{{p_k}}}} \right)^{\frac{1}{{{\alpha _j}}}}}$. Hence, the local delay for a typical user associated to the ${k^{th}}$ tier and operates in FD mode is given by  
\begin{dmath}
	\mathcal{D}_k^{\rm FD} = \int_0^{{\vartheta _k}} {\mathcal{D}_k^{\rm FD}\left( r \right)} .{f_{r_k^{\rm FD}}}\left( r \right)dr,
\end{dmath}

\chr{
\begin{equation}\label{eq: final D_k^FD}
\begin{split}
\mathcal{D}_k^{\rm FD} = \int_0^{{\vartheta _k}}& {\frac{{2\pi {\lambda _k}}}{{\left( {1 - {\chi_{_k}}} \right)\mathcal{A}_k^{\rm FD}}}\exp \left( {s\beta {p_{{u_o}}}} \right)} {{\cal F}_1}\left( {r,\lambda_j } \right)\\
&\prod\limits_{j \in \mathcal{K} } {{{\cal F}_2}\left( {r,\lambda_j ,\tau_k } \right){{\cal F}_3}\left( {r,\lambda_u ,\tau_k } \right)} dr.
\end{split}	
\end{equation}
}
Finally, by using Eq.(\ref{eq: final D_k^FD}) in Eq.(\ref{eq: define total local delay FD/HD}), we get Eq.(\ref{eq: General Eq local delay FD}). The proof of Eq.(\ref{eq: General Eq local delay HD}) is similar to the same process and we avoid the repetition due to lack of space. \\
\end{IEEEproof}

\subsubsection{Special case}\label{subsub: Special Case}

\chr{
By considering ${\alpha _k} = {\alpha _j} = {\alpha _u} = \alpha $ in Eq.(\ref{eq: General Eq local delay FD}), Eq.(\ref{eq: General Eq local delay HD}), Theorem \ref{theory: Local delay} can be simplified as Eq.(\ref{eq: special case Delay-FD}), Eq.(\ref{eq: special case Delay-HD}) at the page \pageref{eq: special case Delay-FD}.
}

\begin{figure*}[t]  	
	\chr{\begin{IEEEeqnarray}{lLl}\label{eq: special case Delay-FD}
		{\mathcal{D}_{\rm FD}}&= \sum\limits_{k \in \mathcal{K} } {\frac{{2\pi {\lambda _k}}}{{\left( {1 - {\chi_{_k}}} \right)}}} \int_0^{{\vartheta _k}} r. {\exp \left( {\frac{{{\tau _k}r^{{\alpha}}}\beta {p_{{u_o}}}}{{{p_k}}}} \right)}
		 \exp \Bigg(- \pi \sum\limits_{j \in \kappa } {{{\left( {\frac{{{p_j}}}{{{p_k}}}} \right)}^{\frac{2}{\alpha }}}{r^2}}\Big({\lambda _j}\big( {1 - \left( {1 - {\chi _j}} \right)\rho \left( {{\tau _k},\alpha } \right)}\big)
	 - \lambda _{_u}^{FD}\rho '\left( {{\tau _k},\alpha } \right) \Big)\Bigg) {dr},\\
	  \label{eq: special case Delay-HD}
	  {\mathcal{D}_{\rm HD}}&= \sum\limits_{k \in \mathcal{K} } {\frac{{2\pi {\lambda _k}}}{{\left( {1 - {\chi_{_k}}} \right)}}} \int_{{\vartheta _k}}^\infty 
	  r.\exp \Bigg(- \pi \sum\limits_{j \in \kappa } {{{\left( {\frac{{{p_j}}}{{{p_k}}}} \right)}^{\frac{2}{\alpha }}}{r^2}}
	  \Big({\lambda _j}\big( {1 - \left( {1 - {\chi _j}} \right)\rho \left( {{\tau _k},\alpha } \right)}\big)
	 - \lambda _{_u}^{FD}\rho '\left( {{\tau _k},\alpha } \right) \Big)\Bigg) {dr},\nonumber\\
	  & \mathop  = \limits^{(a)} \sum\limits_{k \in \mathcal{K} } {\frac{{{\lambda _k}{{p_k}^{\frac{2}{\alpha }}}}}{{\left( {1 - {\chi_{_k}}} \right)}}}.
	  {\frac{\exp \Bigg(- \pi \sum\limits_{j \in \kappa } {{{\left( {\frac{{{p_j}}}{{{p_k}}}} \right)}^{\frac{2}{\alpha }}}{{{\vartheta _k}}^2}}\Big({\lambda _j}\big( {1 - \left( {1 - {\chi _j}} \right)\rho \left( {{\tau _k},\alpha } \right)}\big)- \lambda _{_u}^{FD}\rho '\left( {{\tau _k},\alpha } \right) \Big)\Bigg)}{{\sum\limits_{j \in \kappa } {{p_j}^{\frac{2}{\alpha }} \Big({\lambda _j}\big( {1 - \left( {1 - {\chi _j}} \right)\rho \left( {{\tau _k},\alpha } \right)}\big)- \lambda _{_u}^{FD}\rho '\left( {{\tau _k},\alpha } \right) \Big)} }}}
	\end{IEEEeqnarray}}
	\hrule
\end{figure*}
\chr{In Eqn. (\ref{eq: special case Delay-FD}), (a) follows taking integral with respect to variable $r$, ${\rho \left( {{\tau _k},\alpha } \right)} = \int_1^\infty \left( {\frac{{{\tau _k}}}{{{\tau _k} + {u^{\frac{{{\alpha}}}{2}}}}}} \right)du$, and ${{\rho '} \left( {{\tau _k},\alpha } \right)} = \int_1^\infty \left( {\frac{{{\tau _k}{p_u}}}{{{\tau _k}{p_u} + {u^{\frac{\alpha}{2}}}{p_j}}}} \right)du$.}

\subsection{Energy Efficiency \rm {(EE)}} \label{sub: EE}
\begin{defn}
	\label{EE Defin}
	For a successful communication between the typical user and its associated BS, EE is defined as the ratio of network throughput (or area spectral efficiency) to the average total power consumption \cite{LocalDelay5}.	
\end{defn}
 The achieved throughput in a single HD/FD link is denoted by $\mathcal{D}_{\dag}^{ - 1}\log \left( {1 + \tau } \right)$ nats/s/Hz. The sum throughput of the hybrid HD/FD network is given by ${\mathcal{T}_\dag } = D_\dag ^{ - 1}\log \left( {1 + \tau } \right)\sum\nolimits_{k \in \mathcal{K}} {\left( {1 - {\chi_{_k}}} \right)\lambda _k^\dag } $, where $\lambda _k^\dag  = \frac{{\mathcal{A}_k^\dag }}{{{\mathcal{A}_k}}}$. In this paper, we obtain the power consumption of each BS in each tier based on a linear approximation of the BS power modeling \cite{powr-consume}. The power consumption of a BS in tier $k$ denoted by $P_k$ and is defined by
 
 \begin{equation}
 {P_k} = 
 \begin{cases}
 {{P_{k0}} + {\Delta _k}{p_k}} { + {\mathbbm{1}_\dag}.  {P_{SIC}}},&{{p_k} > 0},\\
{{P_{{\rm{k,sleep}}}}},&{p_k} = 0
 \end{cases}
 \end{equation}
 
 where the above expression indicates the power expenditure in active mode, which ${{P_{k0}}}$, ${{\Delta _k}}$ is the static power and the slope of power consumption in the ${k^{th}}$ tier, respectively. $P_{SIC}$ represents power consumption for SI cancellation. ${\mathbbm{1}_\dag}$ is the indicator function defined in Eq. (\ref{IndFunc}). ${{P_{{\rm{k,sleep}}}}}$ is the static power consumption in sleep mode \cite{powr-consume}. Conclusively, we define the EE as
 
 \begin{equation} \label{eq: define EE}
 \begin{split}
 \eta  = \frac{{D_\dag ^{ - 1}\log \left( {1 + \tau } \right)\sum\limits_{k \in \mathcal{K}} {\left( {1 - {\chi_{_k}}} \right)\lambda _k^\dag } }}{{\sum\limits_{k \in \mathcal{K}} {\lambda _k^\dag \left[ {\left( {1 - {\chi_{_k}}} \right)\left( {{P_{k0}} + {\Delta _k}{p_k} + {\mathbbm{1}_\dag.}{P_{SIC}}} \right) + {\chi_{_k}}{P_{{\text{k,sleep}}}}} \right]} }},
 \end{split}
 \end{equation}
 where the denominator represents power consumption of all BSs across the entire network. The unit of EE is nats/Joule/Hz. By substituting ${\mathcal{D}_{\rm FD}}$ and ${\mathcal{D}_{\rm HD}}$ from \chr{Eq.(\ref{eq: General Eq local delay FD}) and (\ref{eq: General Eq local delay HD}) in (\ref{eq: define EE})}, the general expression of  $\eta $ can be obtained. Since the expression of EE in Eq. (\ref{eq: define EE}) is not in closed form, we will validate it numerically in the next section. 
 
 \section{Numerical Results} \label{sec:NumResults}
\begin{figure}[h]
 	\begin{minipage}[t]{0.5\linewidth}
 		\centering	
 		\subfloat[]	
 		{\includegraphics[width=1.9in]{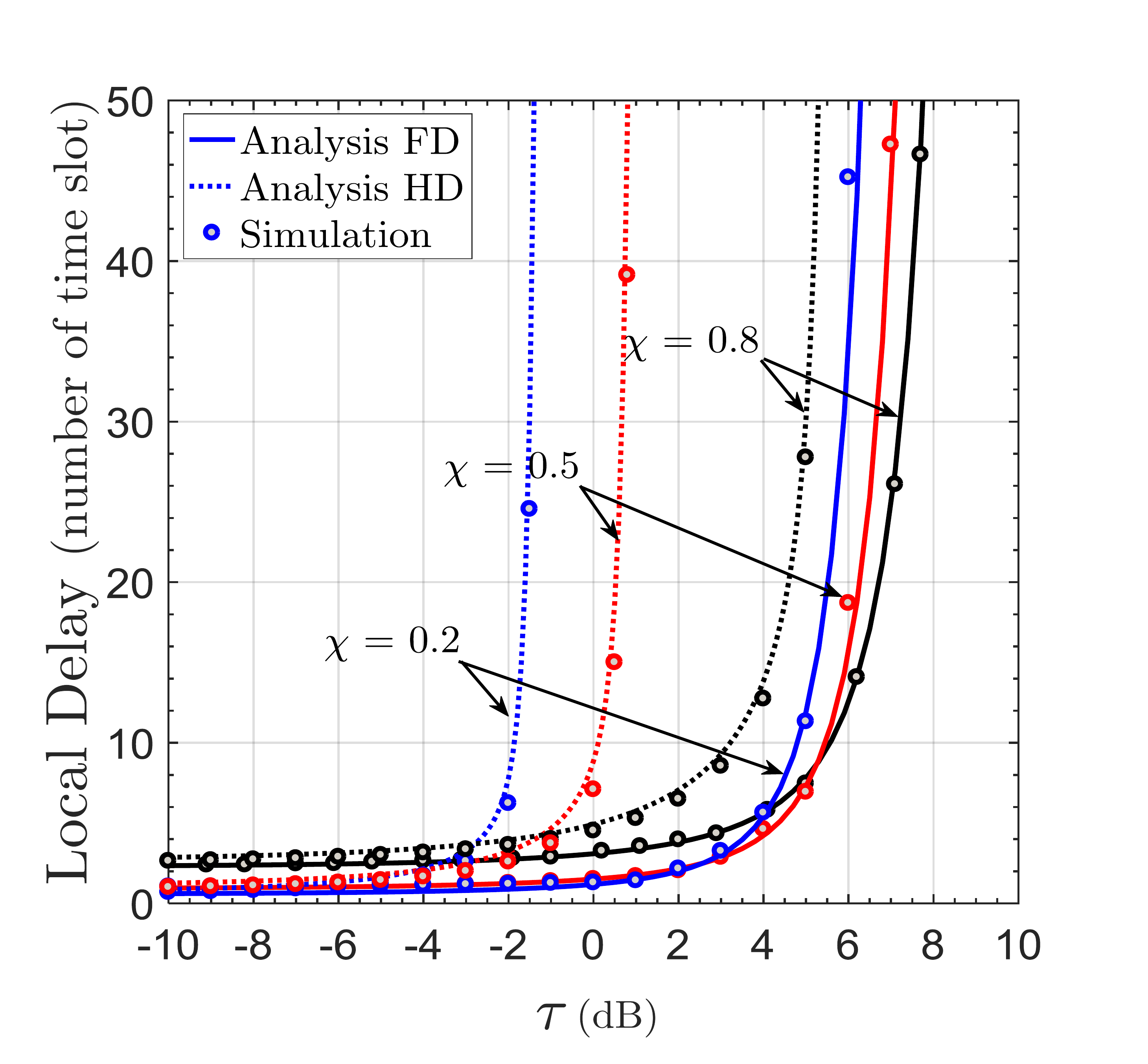}
 			\label{fig: LD vs SIR differ mute}}		
 	\end{minipage}%
 	\begin{minipage}[t]{0.5\linewidth}
 		\centering
 		\subfloat[]			
 		{\includegraphics[width=1.9in]{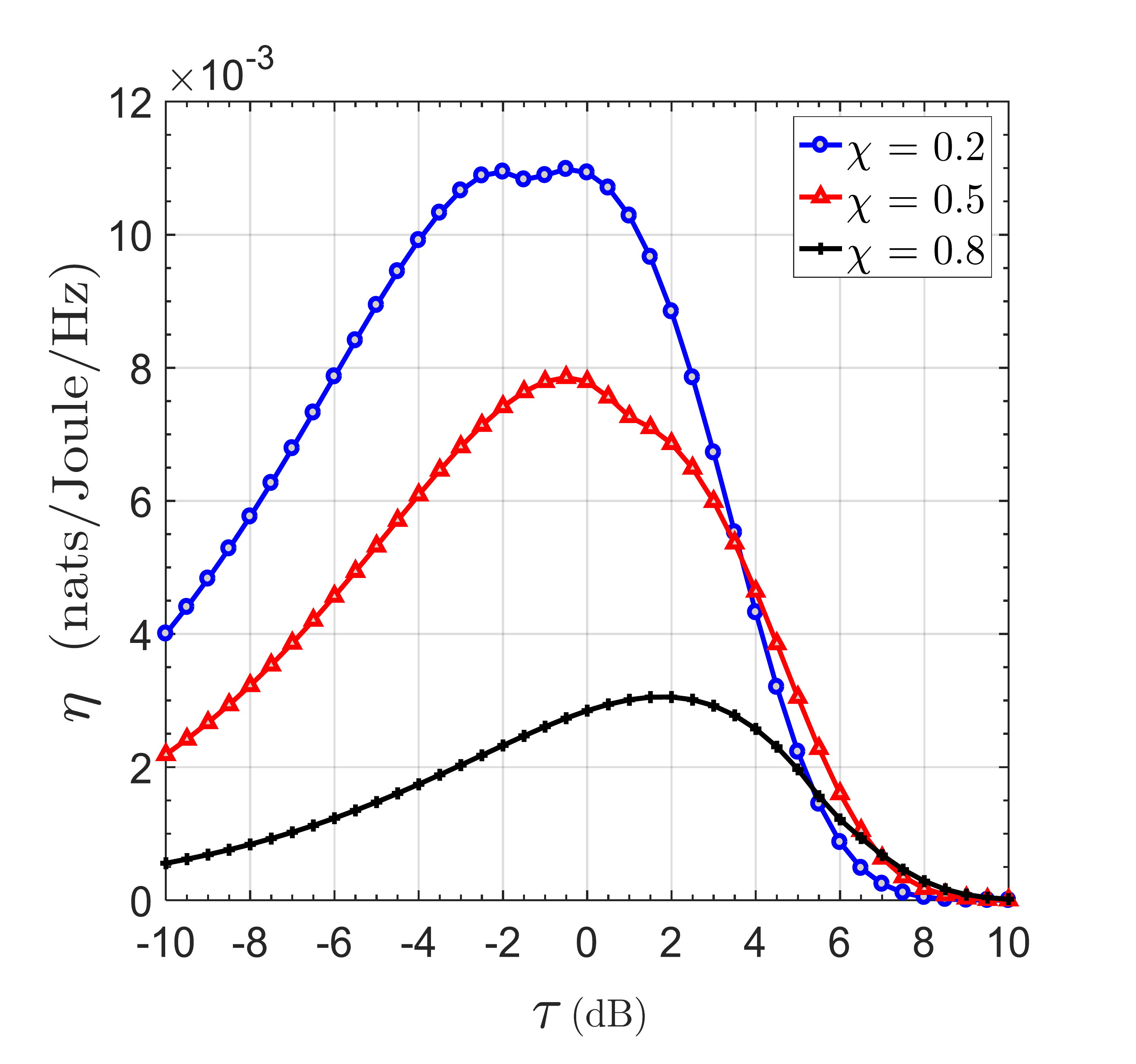}
 			\label{fig: EE vs SIR differ mute}}					
 	\end{minipage}
 	\caption{\scriptsize Local Delay \chr{for a typical user FD and HD in the hybrid network} and EE versus SIR threshold $\tau$ for different values of silent probability, where ${\alpha _1} = {\alpha _2} = {\alpha _u} = 3.5$, ${\lambda _u} = 10{\lambda _2} = 50{\lambda _1} \rm \left( {\frac{{users/BSs}}{{k{m^2}}}} \right)$.}
 	\label{fig: LD-EE vs SIR differ mute}	
 \end{figure}

\begin{figure}[h]	
	\begin{minipage}[t]{0.5\linewidth}
		\centering	
		\subfloat[]	
		{\includegraphics[width=1.9in]{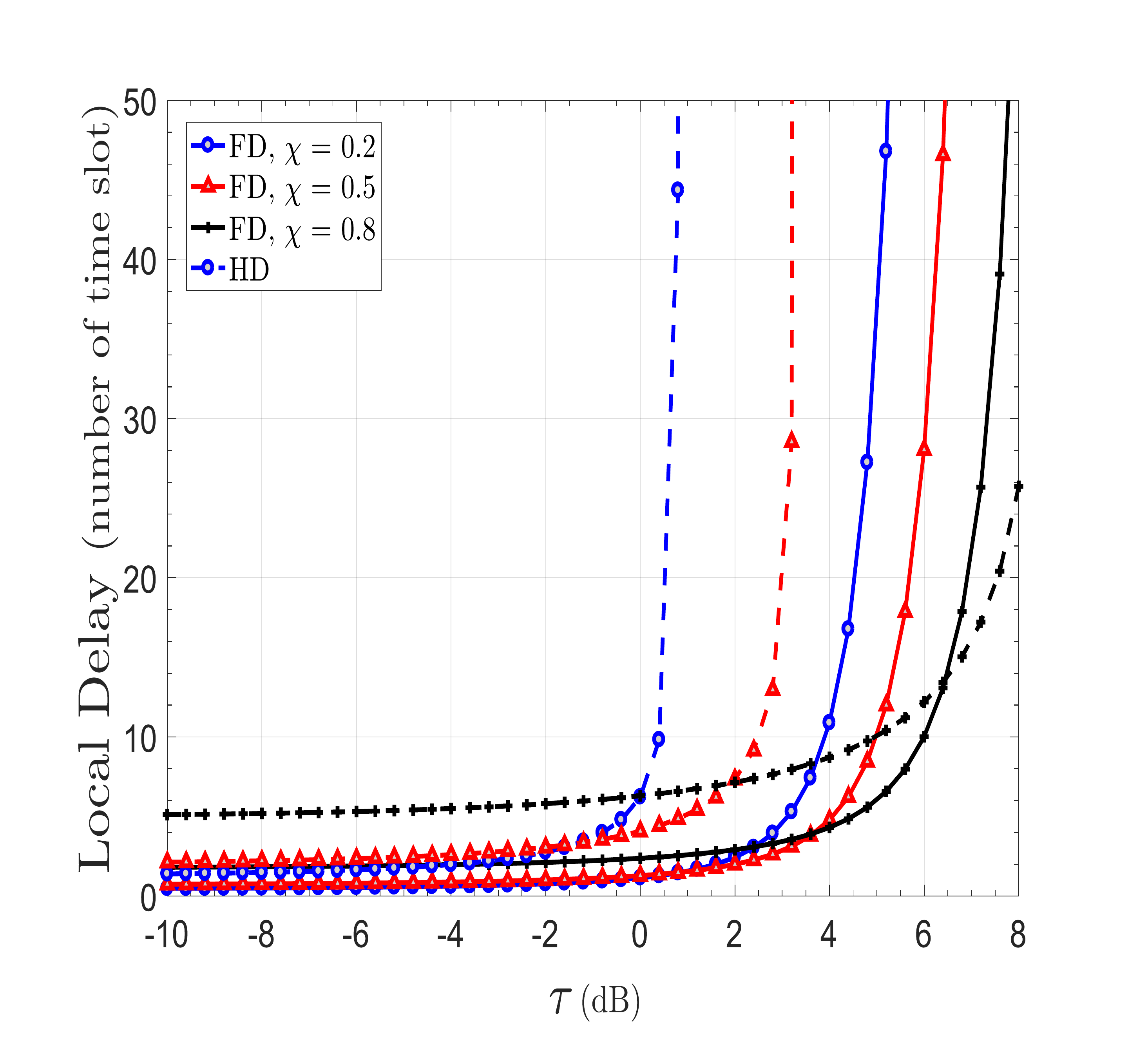}
			\label{fig: LD vs SIR differ mute 3 tier}}		
	\end{minipage}%
	\begin{minipage}[t]{0.5\linewidth}
		\centering
		\subfloat[]			
		{\includegraphics[width=1.9in]{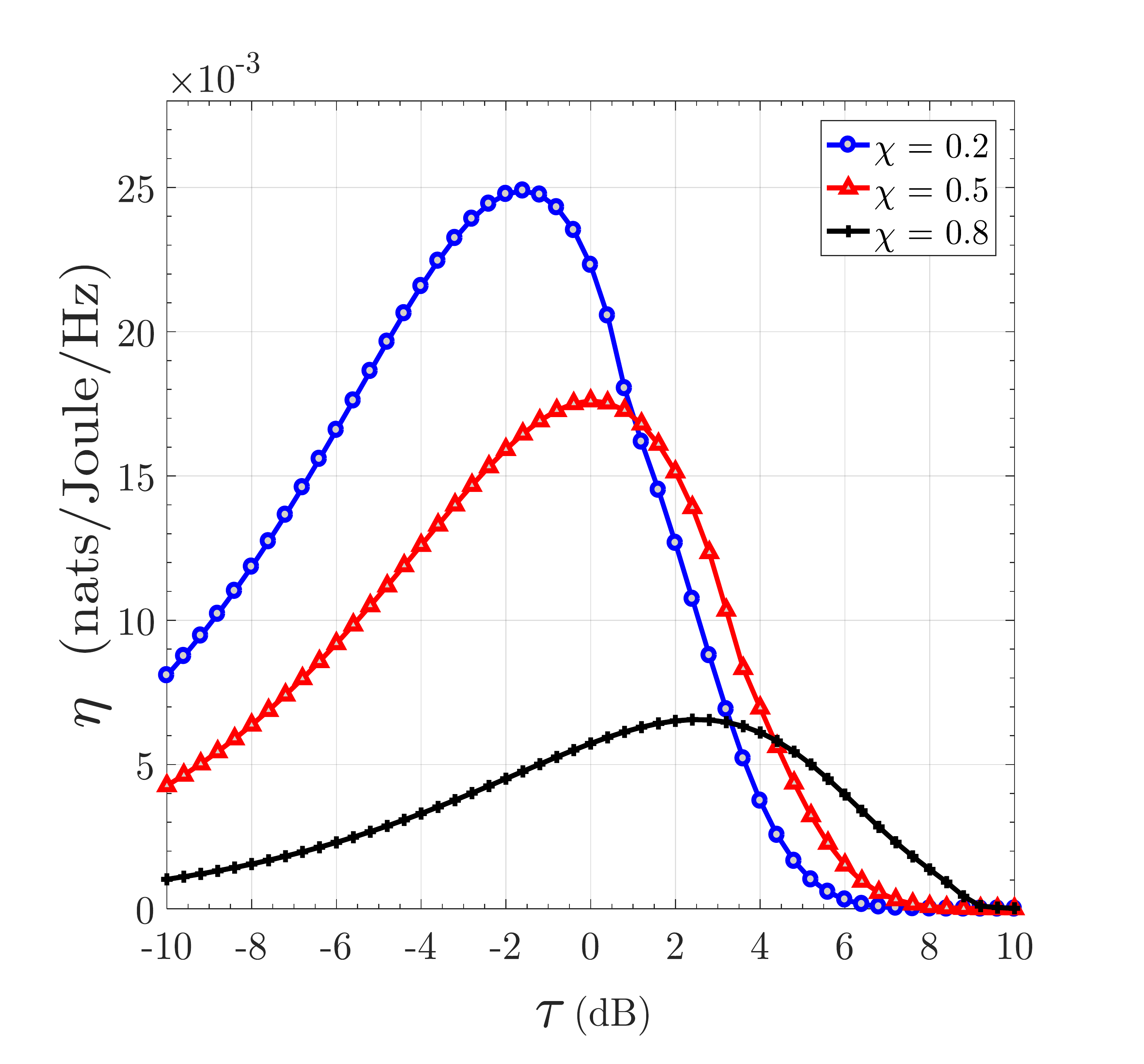}
			\label{fig: EE vs SIR differ mute 3tier}}					
	\end{minipage}
	\caption{\scriptsize Local Delay for a typical user FD and HD in the hybrid network and EE in three-tier hybrid systems versus SIR threshold $\tau$ for different values of silent probability, where ${\alpha _1} = {\alpha _2} = {\alpha _u} = 3.5$, ${\lambda _u} = 50 \rm \left( {\frac{{users}}{{k{m^2}}}} \right)$, ${\lambda _3} = 4{\lambda _2} = 8{\lambda _1} \rm \left( {\frac{{BSs}}{{k{m^2}}}} \right)$.}
	\label{fig: LD-EE vs SIR differ mute 3 tier}	
\end{figure}

\begin{figure}[h!]
 	\begin{minipage}[t]{0.5\linewidth}
 		\centering	
 		\subfloat[]		
 		{\includegraphics[width=1.9in]{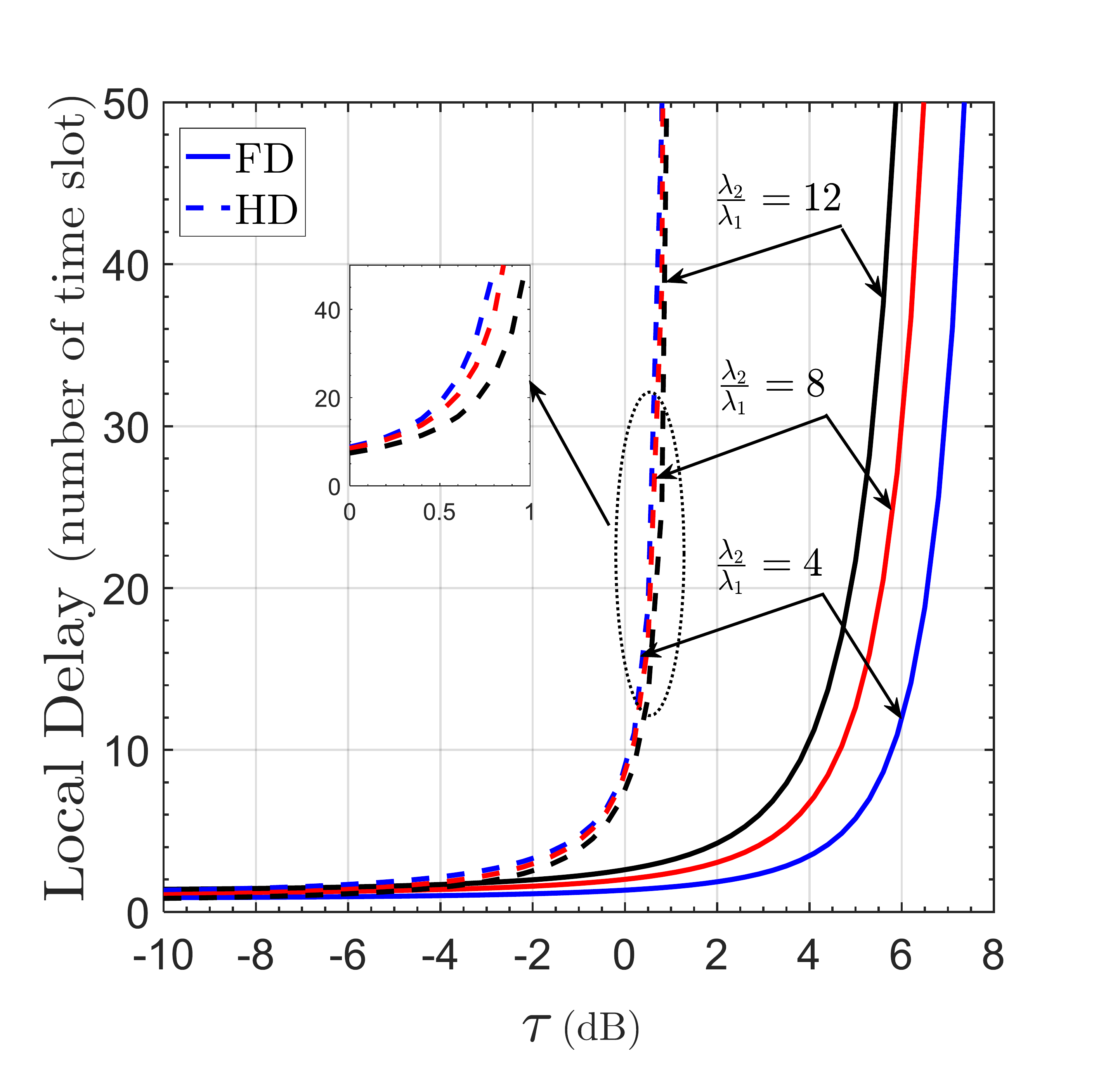}
 			\label{fig: LD vs SIR differ lambda ratio}}			
 	\end{minipage}%
 	\begin{minipage}[t]{0.5\linewidth}
 		\centering	
 		\subfloat[]	
 		{\includegraphics[width=1.9in]{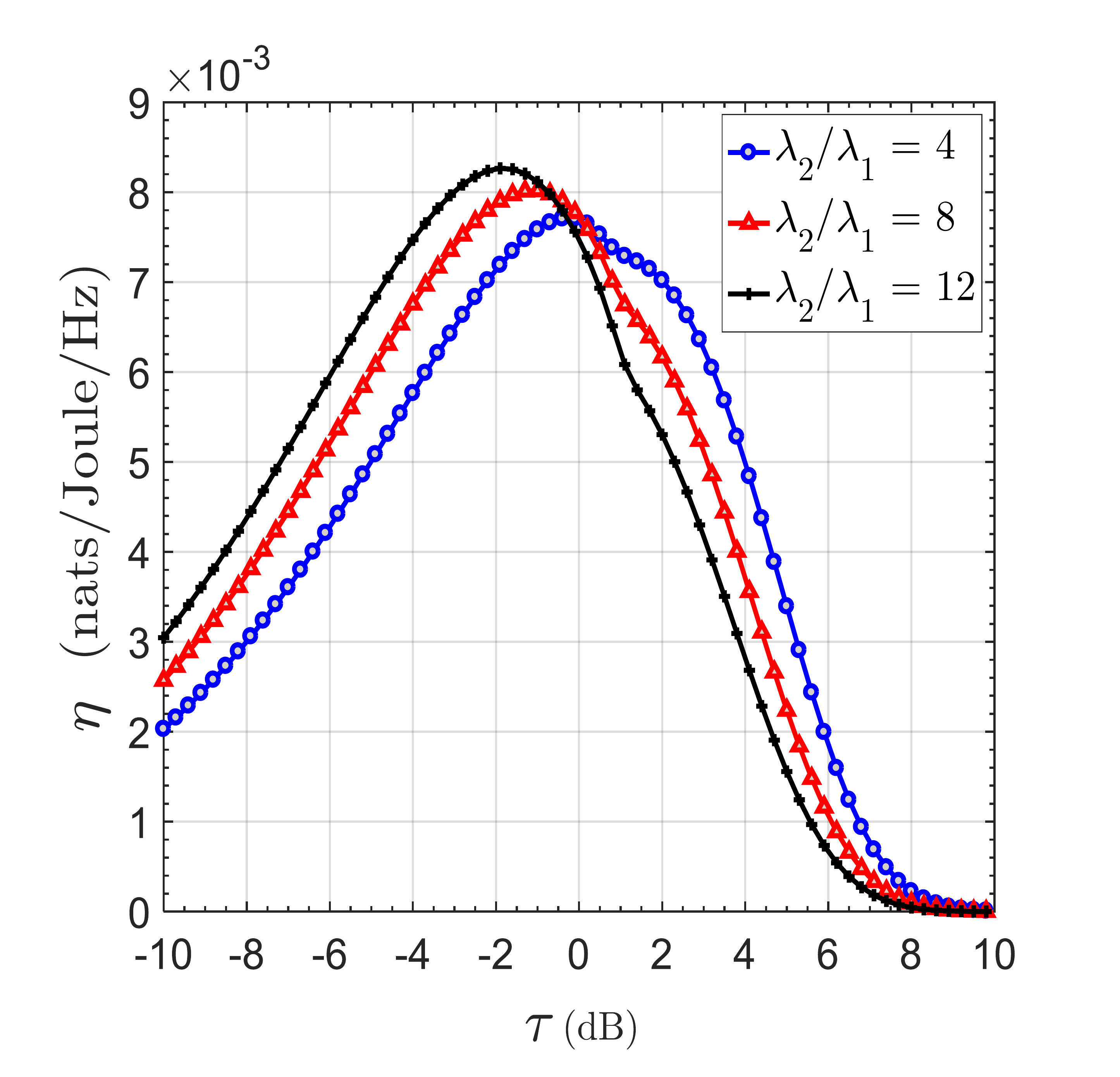}
 			\label{fig: EE vs SIR differ lambda ratio}}			
 	\end{minipage}
 	\caption{\scriptsize Local Delay \chr{for a typical user FD and HD in the hybrid network} and EE versus SIR threshold $\tau$ for different values of $\frac{{{\lambda _2}}}{{{\lambda _1}}}$, where ${\alpha _1} = {\alpha _2} = {\alpha _u} = 3.5$, $\chi _1 = \chi _2 = 0.5$ and ${\lambda _u} = 50  \rm \left( {\frac{{users}}{{k{m^2}}}} \right)$.}
 	\label{fig: LD-EE vs SIR differ lambda ratio}	
 \end{figure}

\begin{figure}[h!]
 	\begin{minipage}[t]{0.5\linewidth}
 		\centering	
 		\subfloat[]			
 		{\includegraphics[width=1.9in]{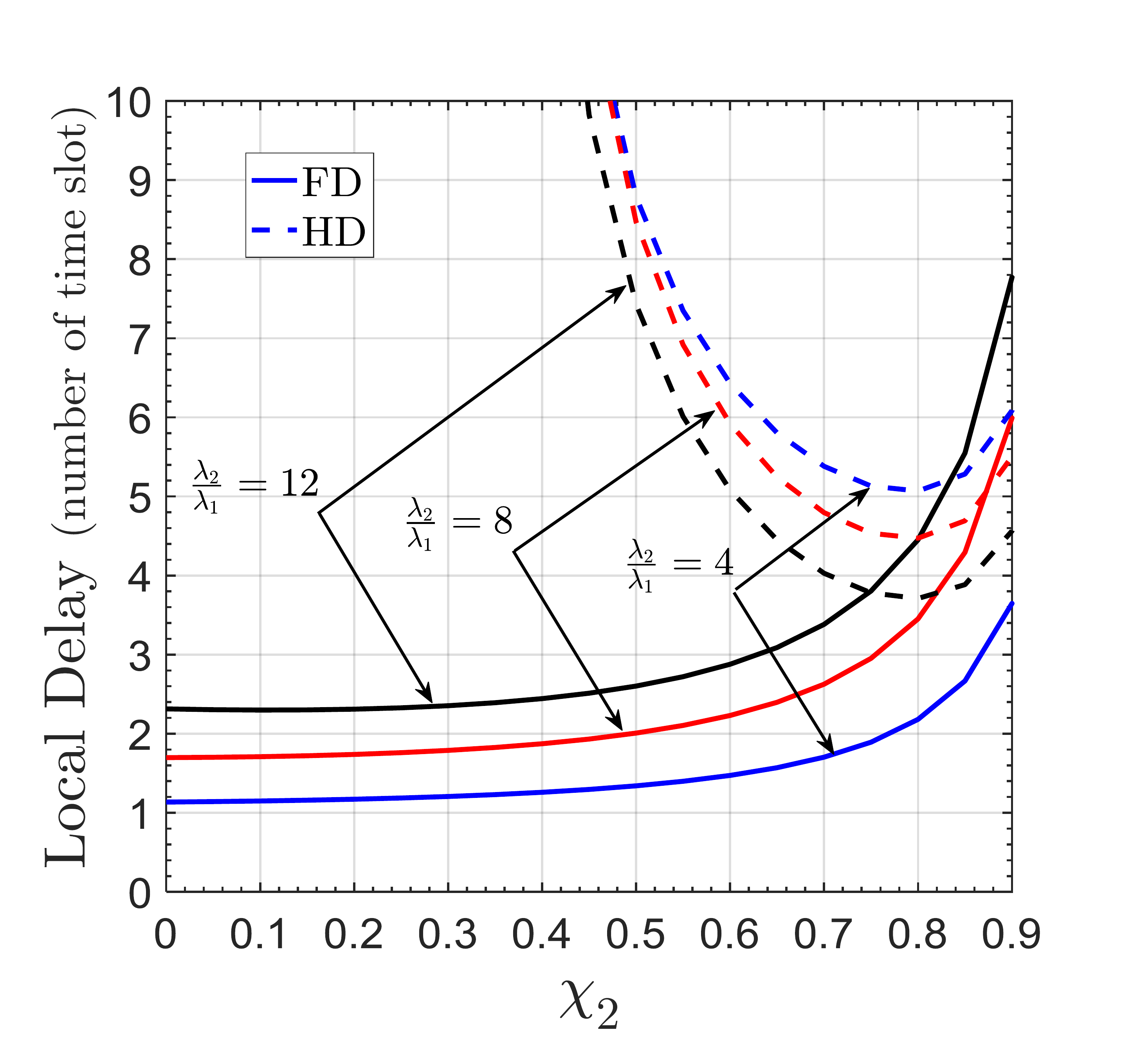}
 			\label{fig: LD vs mute2 for differ lambda2}}		
 	\end{minipage}%
 	\begin{minipage}[t]{0.5\linewidth}
 		\centering		
 		\subfloat[]
 		{\includegraphics[width=1.9in]{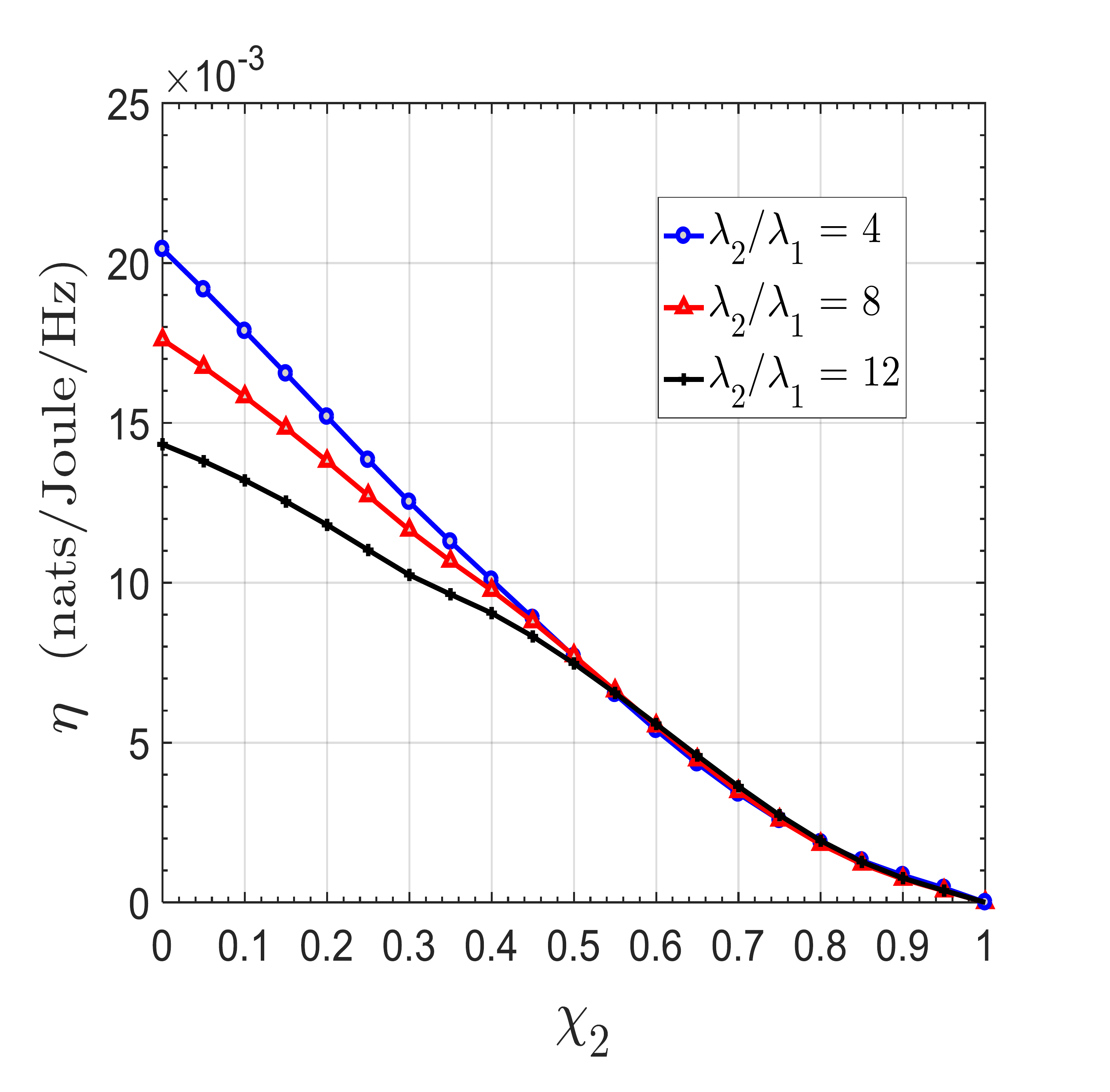}
 			\label{fig: EE vs mute2 for differ lambda2}}		
 	\end{minipage}
 	\caption{\scriptsize Local Delay \chr{for a typical user FD and HD in the hybrid network} and EE versus silent probability (in tier 2) for different values $\frac{{{\lambda _2}}}{{{\lambda _1}}}$, where ${\alpha _1} = {\alpha _2} = {\alpha _u} = 3.5$, $\chi _1 = 0.5$,  $\tau  = 0$ \rm {dB} and ${\lambda _u} = 50 \rm \left( {\frac{{users}}{{k{m^2}}}} \right)$.}
 	\label{fig: LD-EE vs mute2 for differ lambda2}	
 \end{figure}

\begin{figure}[h]  
 	\centering
 	\includegraphics[width=0.26
 	\textwidth]{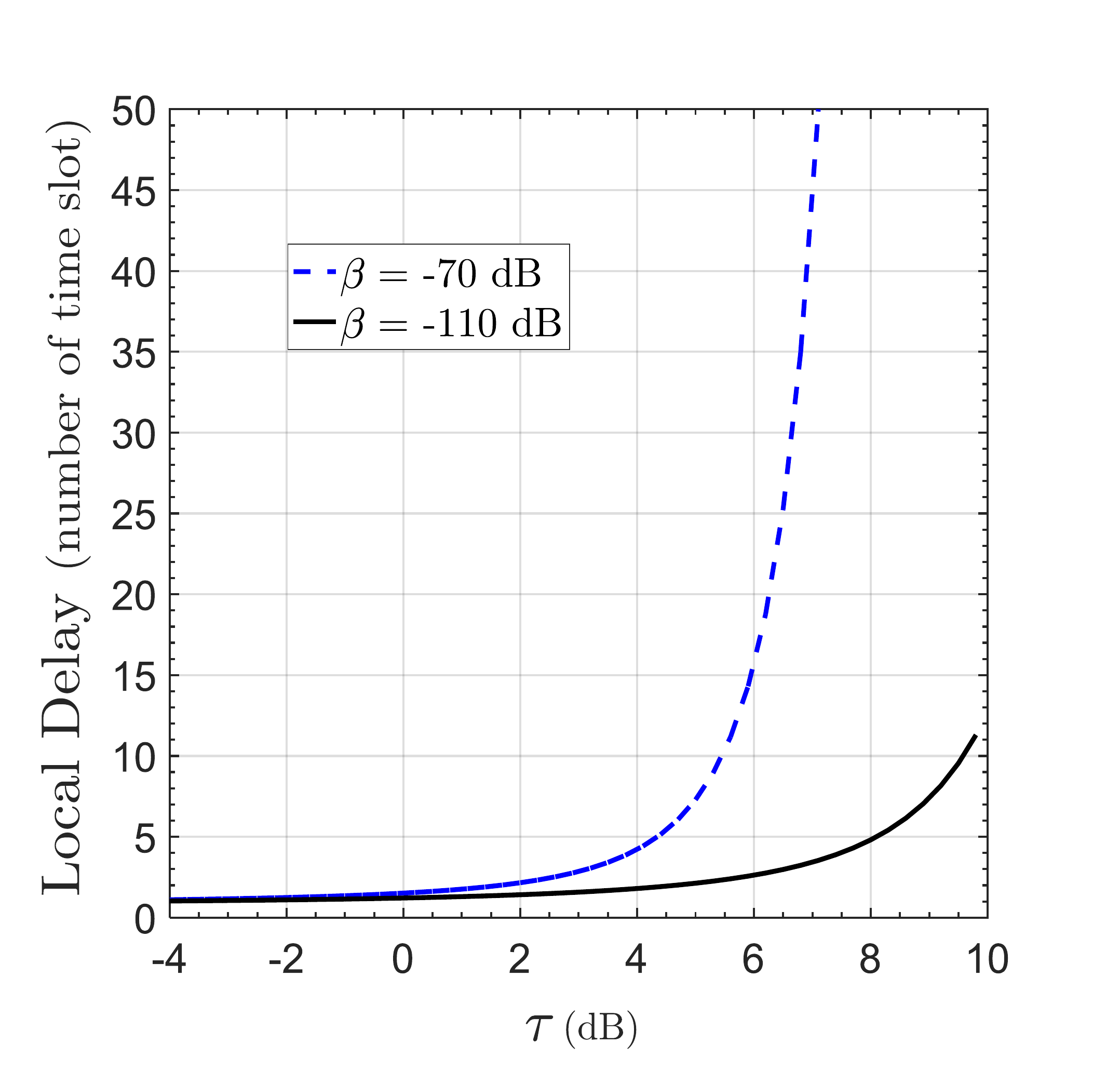}
 	\caption{\scriptsize Local Delay \chr{for a typical user FD in the hybrid network} versus SIR threshold $\tau$ for different values of $ \beta $, where ${\alpha _1} = {\alpha _2} = {\alpha _u} = 3.5$, $\chi _1 = \chi _2 = 0.5$, ${\lambda _u} = 10{\lambda _2} = 50{\lambda _1} \rm \left( {\frac{{users/BSs}}{{k{m^2}}}} \right)$.}
 	\label{fig: LD vs SIR differ beta}
 \end{figure}

\begin{figure}[h!]  
 	\centering
 	\includegraphics[width=0.26
 	\textwidth]{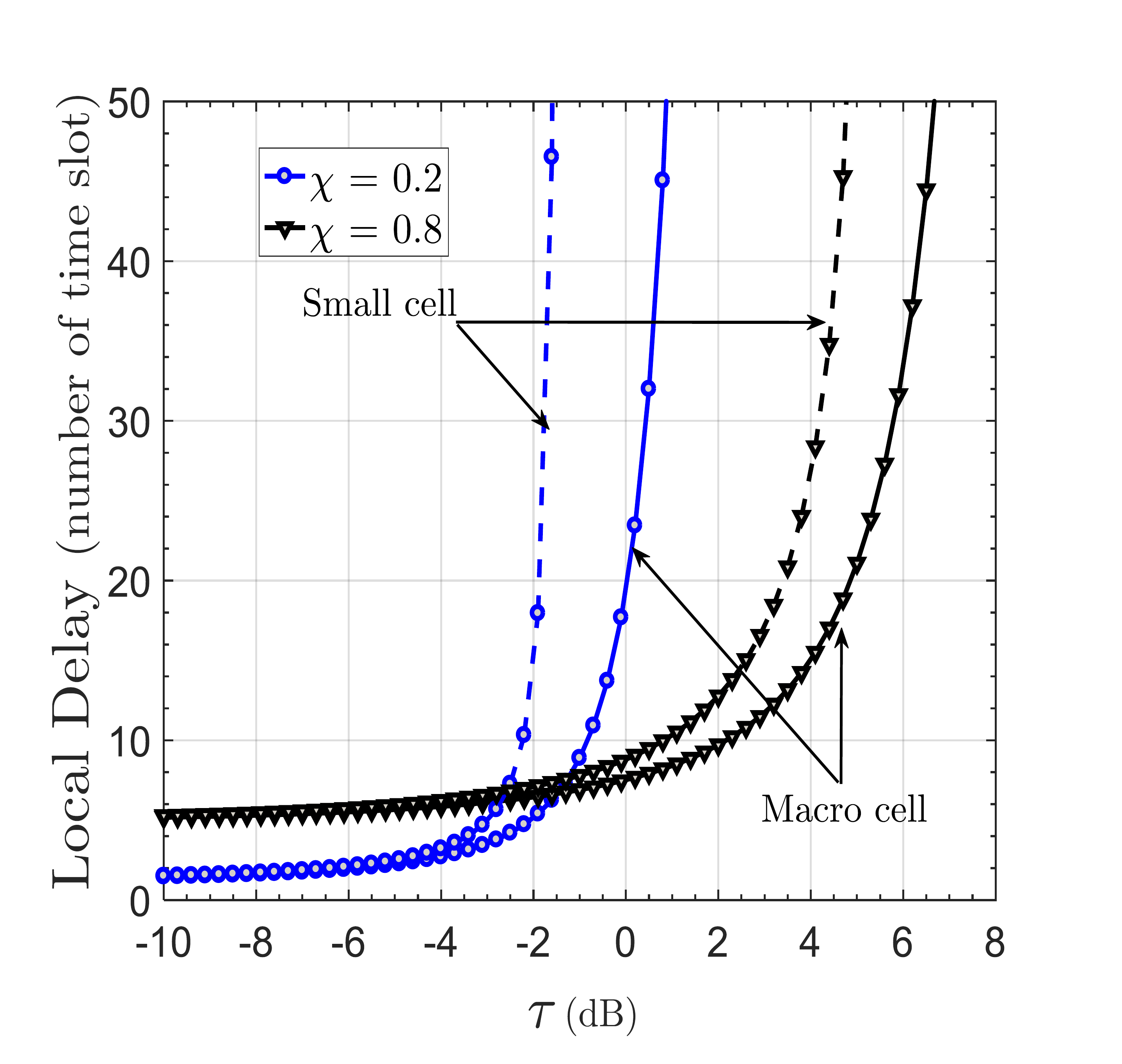}
 	\caption{\scriptsize Local Delay in two tiers versus SIR threshold $\tau$ for different silent probabilities, where ${\alpha _1} = {\alpha _2} = {\alpha _u} = 3.5$, ${\lambda _u} = 10{\lambda _2} = 50{\lambda _1} \rm \left( {\frac{{users/BSs}}{{k{m^2}}}} \right)$.}
 	\label{fig: LD tiers vs SIR }
 \end{figure}

 We consider a two-tier (macro and poico BSs) cellular HetNet ($K=2$). The transmit power for the BSs in first and second tiers assumed to be 46 dBm, 30 dBm, respectively. Transmit power for the user devices is 23 dBm.  ${\vartheta _1} = 300$ m and ${\vartheta _2} = 150$ m. SIC factor is $\beta = -70$ dB and power expenditures are ${P_{10}} = 139$ Watts, ${P_{20}} = 9.7$ Watts, ${\Delta _1} = 5$, ${\Delta _2} = 5.7$, ${P_{SIC}} = 50$ mWatts, ${P_{{\rm{1,sleep}}}} = {\rm{80}}$ Watts and ${P_{{\rm{2,sleep}}}} = {\rm{6.1}}$ Watts. Note that in these parameters, "1", "2" refer to the first and second tier, respectively. Fig. \ref{fig: LD-EE vs SIR differ mute} shows the local delay \chr{for a typical user FD and HD in the hybrid network} and EE versus SIR threshold $\tau$ for different silent probabilities. In Fig. \ref{fig: LD vs SIR differ mute} we can observe that the local delay for FD mode in all silence probabilities is less than the HD one. This is because the FD users are closer to their tagged BSs which leads to much stronger received power by the FD transceivers in comparison with the HD one. However, with the rising of SIR threshold, the local delay for both duplex modes is increasing. We find that in the plane of $\tau  < 0$, by increasing the silent probability, local delay in both HD and FD communications increases. The reason is, at the low regime of $\tau$, most of the transmissions are successful and consequently the interference from BSs on the typical user has slight impact. Conversely, in the high regime of $\tau$ (i.e., $\tau > 0$), the impact of aggregated interference is more accented. However, by increasing the silent probability, the number of active BSs decreases, which will lead to partial reduction in the suffered interference at the typical user and this will result in local delay reduction. As can be seen from Fig.\ref{fig: EE vs SIR differ mute}, EE is being reduced by increasing the silent probability. It is clear that by increasing the silent probability, the number of active BSs decreases, namely the number of active links between user devices and BSs will be reduced. This causes reduction in system sum throughput, while the BSs in silent mode use energy, correspondingly, the EE will be decreased according to its definition in \ref{EE Defin}. It can be seen that there is an optimal SIR threshold (around 0 dB) that maximizes EE. \chr{In Fig. \ref{fig: LD-EE vs SIR differ mute 3 tier}, local delay \chr{for a typical user FD and HD in the hybrid network} and EE are illustrated for a three-tier HetNet, in which the transmission power for the third tier assumed to be 23 dBm, ${\vartheta_3} = 75$ m, and power expenditures for the third tier are ${P_{30}} = 7.7$ Watts, ${\Delta _3} = 7$, and ${P_{{\rm{3,sleep}}}} = {\rm{4.8}}$ Watts. By increasing the number of tiers, due to the reduction in distance threshold, the chance of making FD communications, decreases. This is due to the fact that in the third tier, the distance threshold is much smaller than the distance thresholds in tiers one and two. Hence, the existing HD users in tire three experience lower delay due to the reduction in the interference incurred from the FD users, therefore they can receive higher power levels. Correspondingly, this will incur three-tier HetNet a lower delay, higher throughput, and consequently a better energy efficiency in comparison to two-tier HetNet. The explanation of EE in Fig. \ref{fig: EE vs SIR differ mute} is also valid for Fig. \ref{fig: EE vs SIR differ mute 3tier}. The effect of silent probability in three-tier HetNet is similar to the results of two-tier network demonstrated in Fig. \ref{fig: LD-EE vs SIR differ mute}.} In Fig.\ref{fig: LD vs SIR differ lambda ratio}, by increasing the density of tier 2, the probability that more users can operate in FD mode is raising. On the other hand, interference from FD users on a typical user increases and due to distance-based duplex selection scheme, the FD users are getting closer to each other. This phenomena causes higher aggregate interferences on the typical user and consequently will increase the chance of link failure. As a result, the local delay of the typical user that operates in FD mode will be increased. Conversely, for the HD counterpart, by increasing the density of users across the entire network, fewer users will operate in HD mode (i.e., due to distance-based duplex selection scheme), and therefore these users will be located at the far distant away. Correspondingly, the impact of aggregated interference from FD users on the typical user that operates in HD mode will be decreased. Hence, the local delay will decrease or not change with higher density. Conceding to Fig. \ref{fig: EE vs SIR differ lambda ratio}, in the higher SIR threshold regime ($\tau \ge {\rm{0}}$), increasing density causes more power consumption. On the other hand, due to non-satisfaction of the SIR threshold, we will have increment in local delay and reduction in EE. However, in the lower SIR thresholds ($\tau \le 0$), the impact of short distance association dominates against the aggregated interference, and hence the acquired SIR is improved and EE is enhanced. \chr{Fig. 5a demonstrates the local delay as a function of the silent probability of BSs of tier 2 (i.e., $\chi_2$). By increasing the chance of being silent, the probability of operating in FD mode and consequently the number of FD transceivers decreases due to the reduction in the number of BSs in tier 2. Hence, the aggregated interference imposed on typical HD user decreases and as a result, the local delay decreases. On the other hand, reduction in the number of active BSs causes reduction in the success probability for the users, which in turn leads to higher local delay.} There is an optimal silence probability to minimize local delay when a typical user operates in HD mode. In case of the impact of density, similar explanations for Fig. \ref{fig: LD vs SIR differ lambda ratio} are also valid for Fig. \ref{fig: LD vs mute2 for differ lambda2}. For the EE as shown in Fig.\ref{fig: EE vs mute2 for differ lambda2}, with respect to Fig. \ref{fig: EE vs SIR differ lambda ratio}, by increasing density of BSs, the EE decreased at 0 dB of $\tau$. It is due to the increasing aggregated interference, high energy consumption, non-satisfaction of SIR threshold and reduction in system sum throughput. As can be seen, the curves overlap with increasing the chance of being silence in tier 2 due to reduction in the effective interference from tier 2. As a result, the throughput and consequently EE is increased. The local delay \chr{for a typical user FD in the hybrid network} of DL channel with the different SIC factors is shown in Fig.\ref{fig: LD vs SIR differ beta}. As the value of SIC increases, local delay decreases. Obviously, this is due to the better SIC. In Fig.\ref{fig: LD tiers vs SIR } the local delay in DL of macro and small cell users is defined as follows
  \begin{equation}
 {\mathcal{D}_k} = \frac{1}{{{\mathcal{A}_k}}}\left( {\mathcal{A}_k^{\rm FD}\mathcal{D}_k^{\rm FD} + \mathcal{A}_k^{\rm HD}\mathcal{D}_k^{\rm HD}} \right).
 \end{equation}
 For $-10 \le \tau \le -2 $, there is no significant difference in local delay between macro and small cells. However, for the higher SIR thresholds, the local delay in macro cells is less than small cells, and this is because of transmitting power of macro cell which is more than small cells (i.e. Femto and Pico cells) and eventually satisfies the SIR threshold.
 
 \section{Conclusion} \label{sec:conclusion} 
 We analyzed the local delay of $K$-tier HetNet with hybrid HD/FD communications. By using tools from the stochastic geometry, we derived the local delay expression for both HD and FD modes. Analytical and simulation results gave some insights on the impact of key system parameters, including the SIR threshold, BS density and silent probability. Higher SIR threshold increases the local delay. However, from the energy efficiency (EE) point of view, we have optimal limit that increases the network EE. Higher density of the BSs in the high SIR regime causes higher local delay for FD users as well as inefficiency in the energy consumption. Rising the silent probability in the high SIR regime can improve local delay and EE. We observed that Hybrid HetNets can increase the optimal SIR range for EE. Exploring delay-constrained mode selection will be a useful extension to this work. Moreover, in this paper and all related works, the channel for the FD links is assumed to be reciprocal and exchanging data on FD nodes is assumed to be symmetric. FD communications with asymmetric data and a dissimilar channel is another worthwhile open problem to pursue.
 
 \vspace*{-7mm}
 \appendix
 \subsection {Proof for Lemma \ref{lemma: association Prob.}}
 \label{appendix-Associate}
\chr{The event of ${{r_k} \le {\vartheta _k}}$ for FD users, the association probability with the $k$th tier can be given as } 
\chr{
	\begin{equation}\label{eq: Appendix-associate}
				\begin{split}
		&{\mathcal{A}_k^{\rm FD}}  =  {\mathbb{E}_{{r_k}}}\left[{\mathop \prod \limits_{i \ne k} }{\mathcal{P}}\left( { {{r_k} < {{\left( {\frac{{{p_k}}}{{{p_i}}}r_i^{{\alpha _i}}} \right)}^{\frac{1}{{{\alpha _k}}}}}} } \right) \wedge \left( {{r_k} \le {\vartheta _k}} \right) \right] \\
		& = \int_0^{{\vartheta _k}} {\mathop \prod \limits_{i \ne k} {\mathcal{P}}\left({{r_i} > {{\left( {\frac{{{p_i}}}{{{p_k}}}{x^{{\alpha _k}}}} \right)}^{\frac{1}{{{\alpha _i}}}}}} \right) .{f_{{r_k}}}\left( x \right)dx} \\
		& \mathop  = \limits^{\left( a \right)} 2\pi {\lambda _k}\int_0^{{\vartheta _k}} {x.\exp \left( { - \pi \sum\limits_{i \ne k} {{\lambda _i}{{\left( {\frac{{{p_i}}}{{{p_k}}}{x^{{\alpha _k}}}} \right)}^{\frac{2}{{{\alpha _i}}}}} - \pi {\lambda _k}{x^2}} } \right)dx} 
				\end{split}
	\end{equation}
	} 
 
 \chr{where (a) derived by using null probability and distance distribution: $\mathcal{P}\left({{r_i} > {{\left( {\frac{{{p_i}}}{{{p_k}}}{x^{{\alpha _k}}}} \right)}^{\frac{1}{{{\alpha _i}}}}}} \right)= \exp \left( { - \pi {\lambda _i}{{\left( {\frac{{{p_i}}}{{{p_k}}}{x^{{\alpha _k}}}} \right)}^{\frac{2}{{{\alpha _i}}}}}} \right)$ and ${f_{{r_k}}}\left( x \right) = 2\pi {\lambda _k}x\exp \left( { - \pi {\lambda _k}{x^2}} \right)$. By combining both equations, we obtain Eqn. (\ref{eq: Appendix-associate})}. The proof for HD mode is the same.  

\ifCLASSOPTIONcaptionsoff
  \newpage
\fi

\bibliographystyle{IEEEtran}

\begin{thebibliography}{10}
	\bibitem{HetNet2}
	A. Ghosh, N. Mangalvedhe, R. Ratasuk and \textit{et al.}, ``Heterogeneous cellular networks: From theory to practice," \textit{IEEE Commun. Mag.}, vol. 50, no. 6, pp. 54-64, Jun. 2012.
	
	\bibitem{HD}
	A. Gohil, H. Modi and S. K. Patel, ``{5G} technology of mobile communication: A survey," in \textit{Proc., ISSP.}, Mar. 2013, pp. 288-292.
	
	\bibitem{FD1}
	S. Goyal, P. Liu, S. S. Panwar and \textit{et al.}, ``Full duplex cellular systems: will doubling interference prevent doubling capacity?," \textit{IEEE Commun. Mag.}, vol. 53, no. 5, pp. 121-127, May. 2015.
	
	\bibitem{FD2}
	G. Noh, H. Wang, C. Shin and \textit{et al.}, ``Enabling technologies toward fully {LTE}-compatible full-duplex radio," \textit{IEEE Commun. Mag.}, vol. 55, no. 3, pp. 188-195, Mar. 2017.
	
	\bibitem{FD-app1}
	M. Naslcheraghi, S. A. Ghorashi and M. Shikh-Bahaei, ``{FD} device-to-device communication for wireless video distribution," \textit{IET Commun.}, vol. 11, no. 7, pp. 1074-1081,  2017.
	
	\bibitem{FD-app2}
	M. Naslcheraghi, S. A. Ghorashi and M. Shikh-Bahaei, ``Full-duplex device-to-device collaboration for low-latency wireless video distribution," in \textit{Proc., ICT}, May. 2017, pp. 1-5.
	
	\bibitem{FD-app3}
	M. Naslcheraghi, M. Afshang and H. S. Dhillon, ``Modeling and performance analysis of full-Duplex communications in cache-enabled {D2D} Networks," in \textit{Proc., IEEE  ICC}, May. 2018, pp. 1-6.
	
	\bibitem{FDSurvey}
	A. H. Gazestani, S. A. Ghorashi, B. Mousavinasab and M. Shikh-Bahaei, ``A survey on implementation and applications of full duplex wireless communications," \textit{Physical Commun.}, vol. 34, pp. 121-134, Jun. 2019.
	
	\bibitem{LocalDelay1}
	M. Haenggi, ``The local delay in {P}oisson networks," \textit{IEEE Trans. Inf. Theory}, vol. 59, no. 3, pp. 1788-1802, Mar. 2013.
	
	\bibitem{EditorPaper}
	L. Dai, B. Wang, Z. Ding and \textit{et al.}, ``A survey of non-orthogonal multiple access for 5{G}," \textit{IEEE Commun. Surveys Tuts.}, vol. 20, no. 3, pp. 2294-2323, May. 2018.
	
	\bibitem{Delay-cause}
	Y. Zhong, M. Haenggi, F. Zheng and \textit{et al.}, ``Toward a tractable delay analysis in ultra-dense networks," \textit{IEEE Commun. Mag.}, vol. 55, no. 12, pp. 103-109, Dec. 2017.
	
	\bibitem{LocalDelay2}
	M. Haenggi, ``Local delay in {P}oisson networks with and without interference," in \textit{Proc., Allerton}, Sep. 2010, pp. 1482-1487.
	
	\bibitem{LocalDelay3}
	Z. Gong and M. Haenggi, ``The local delay in {P}oisson networks," \textit{IEEE Trans. Wireless Commun.}, vol. 12, no. 9, pp. 4766-4777, Sep. 2013.
	
	\bibitem{LocalDelay4}
	G. Alfano, R. Tresch and M. Guillaud, ``Spatial diversity impact on the local delay of homogeneous and clustered wireless networks," in \textit{Proc., WSA}, Feb. 2011, pp. 1-6.
	
	\bibitem{Reviewers}
	Y. Zhong, X. Ge, H. H. Yang and \textit{et al.}, ``Traffic matching in 5{G} ultra-dense networks," \textit{IEEE  Commun. Mag.}, vol. 56, no. 8, pp. 100-105, Aug. 2018.
	
	\bibitem{Reviewers2}
	Y. Zhong, X. Ge, T. Han and  \textit{et al.}, ``Tradeoff between delay and physical layer security in wireless networks," \textit{IEEE J. Sel. Areas Commun. }, vol. 36, no. 7, pp. 1635-1647, Jul. 2018.
	
	\bibitem{Hybrid-dulex1}
	J. Lee, T. Q. S. Quek, ``Hybrid full-/half-duplex system analysis in heterogeneous wireless networks," \textit{IEEE Trans. Wireless Commun.}, vol. 14, no. 5, pp. 2883-2895, May. 2015.
	
	\bibitem{Hybrid-duplex2}
	W. Tang and \textit{et al.}, ``Distance-based hybrid duplex in heterogeneous networks," in \textit{Proc., IEEE GLOBECOM}, Dec. 2015, pp. 1-6.
	
	\bibitem{Hybrid-duplex3}
	W. Tang, S. Feng, Y. Liu and Y. Ding, ``Hybrid duplex switching in heterogeneous networks," \textit{IEEE Trans. Wireless Commun.}, vol. 15, no. 11, pp. 7419-7431, Nov. 2016.
	
	\bibitem{LocalDelay5}
	L. Liu, Y. Zhong, W. Zhang and M. Haenggi, ``On the impact of coordination on local delay and energy efficiency in {P}oisson networks," \textit{IEEE Wireless Commun. Lett.}, vol. 4, no. 3, pp. 241-244, Jun. 2015.
	
	\bibitem{DTX-LocalDelay}
	W. Nie, Y. Zhong, F. Zheng and \textit{et al.}, ``Het{N}ets with random {DTX} scheme: Local delay and energy efficiency," \textit{IEEE Trans. Veh. Technol.}, vol. 65, no. 8, pp. 6601-6613, Aug. 2016.
	
	\bibitem{EE-HD}
	D. W. K. Ng and \textit{et al}, ``Energy-efficient resource allocation in {OFDMA} systems with large numbers of base station antennas," \textit{IEEE Trans. Wireless Commun.}, vol. 11, no. 9, pp. 3292-3304, Sep. 2012.
	
	\bibitem{distance-criteria}
	K. T. Hemachandra and A. O. Fapojuwo, ``Distance based duplex mode selection in large scale peer-to-peer wireless networks," in \textit{Proc., IEEE ICC}, May. 2017, pp. 1-6.
	
	\bibitem{EE-Hybrid}
	Z. Wei, S. Sun, X. Zhu and \textit{et al.}, ``Energy-efficient hybrid duplexing strategy for bidirectional distributed antenna systems," \textit{IEEE Trans. Veh. Technol.}, vol. 67, no. 6, pp. 5096-5110, Jun. 2018.
	
	\bibitem{typicaluser-Palm}
	M. Haenggi, \textit{Stochastic Geometry for Wireless Networks}, 2013.
	
	\bibitem{LocalDelay6}
	X. Zhang and M. Haenggi, ``Delay-optimal power control policies," \textit{IEEE Trans. Wireless Commun.}, vol. 11, no. 10, pp. 3518-3527, Oct. 2012.
	
	\bibitem{Stochastic-HetNet2}
	H. Jo, Y. J. Sang, P. Xia and J. G. Andrews, ``Heterogeneous cellular networks with flexible cell association: A comprehensive downlink {SINR} analysis," \textit{IEEE Trans. Wireless Commun.}, vol. 11, no. 10, pp. 3484-3495, Oct. 2012.
	
	\bibitem{powr-consume}
	G. Auer, V. Giannini, C. Desset and \textit{et al.}, ``How much energy is needed to run a wireless network?," \textit{IEEE Wireless Commun. Mag.}, vol. 18, no. 5, pp. 40-49, Oct. 2011.
\end{thebibliography}


\end{document}